\PassOptionsToPackage{table}{xcolor}
\documentclass[10pt,twocolumn,letterpaper]{article}

\usepackage{iccv}              

\usepackage{graphicx}
\usepackage{subcaption}
\usepackage[export]{adjustbox}
\usepackage[table]{xcolor}

\usepackage{multirow}

\usepackage{algorithm}
\usepackage{algorithmic}
\usepackage{amsmath}
\usepackage{amssymb}

\definecolor{iccvblue}{rgb}{0.21,0.49,0.74}
\usepackage[pagebackref,breaklinks,colorlinks,allcolors=green]{hyperref}
\hypersetup{
    citecolor=green,       
    linkcolor=red,       
    urlcolor=blue          
}

\def\paperID{9} 
\def\confName{ICCV}
\def\confYear{2025}

\title{UGPL: Uncertainty-Guided Progressive Learning for Evidence-Based Classification in Computed Tomography}

\author{
    Shravan Venkatraman$^{1}$\thanks{Equal contribution.} \and
    Pavan Kumar S$^{1}$\footnotemark[1] \and
    Rakesh Raj Madavan$^{2}$\footnotemark[1] \and
    Chandrakala S$^{2}$
    \vspace{0.01cm}
    \and 
    $^1$Vellore Institute of Technology, Chennai, India \\ 
    $^2$Shiv Nadar University, Chennai, India 
}

\begin{document}
\maketitle
\begin{abstract}
Accurate classification of computed tomography (CT) images is essential for diagnosis and treatment planning, but existing methods often struggle with the subtle and spatially diverse nature of pathological features. Current approaches typically process images uniformly, limiting their ability to detect localized abnormalities that require focused analysis. We introduce UGPL, an \underline{u}ncertainty-\underline{g}uided \underline{p}rogressive \underline{l}earning framework that performs a global-to-local analysis by first identifying regions of diagnostic ambiguity and then conducting detailed examination of these critical areas. Our approach employs evidential deep learning to quantify predictive uncertainty, guiding the extraction of informative patches through a non-maximum suppression mechanism that maintains spatial diversity. This progressive refinement strategy, combined with an adaptive fusion mechanism, enables UGPL to integrate both contextual information and fine-grained details. Experiments across three CT datasets demonstrate that UGPL consistently outperforms state-of-the-art methods, achieving improvements of 3.29\%, 2.46\%, and 8.08\% in accuracy for kidney abnormality, lung cancer, and COVID-19 detection, respectively. Our analysis shows that the uncertainty-guided component provides substantial benefits, with performance dramatically increasing when the full progressive learning pipeline is implemented. Our code is available at \href{https://github.com/shravan-18/UGPL}{\textcolor{magenta}{https://github.com/shravan-18/UGPL}}.
\end{abstract}    
\section{Introduction}
\label{sec:intro}

Medical image classification plays a central role in automated diagnosis and clinical decision support~\cite{dlformedimg1}. Deep learning and convolutional neural networks (CNNs) have shown effectiveness across various imaging modalities, including X-rays~\cite{xray1,xray2}, magnetic resonance imaging~\cite{mri1,mri2}, and computed tomography (CT) scans~\cite{ct1,ct2}. Particularly in CT image analysis, these approaches have achieved promising results for diagnosing pulmonary diseases~\cite{pul1,pul2}, abdominal abnormalities~\cite{abd1,abd2}, and COVID-19 infections~\cite{cov1,cov2}. While these approaches achieve strong performance on benchmark datasets, they operate uniformly across all spatial regions, overlooking how radiologists selectively attend to diagnostically relevant areas. This limitation affects performance in cases where critical findings are localized and subtle.

\begin{figure}[t]
    \centering

    \begin{subfigure}[t]{0.48\columnwidth}
        \centering
        \includegraphics[width=\linewidth]{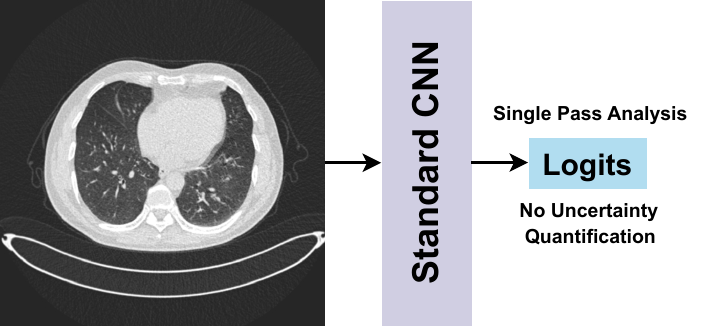}
        \caption{Standard CNN: Single-pass analysis with uniform processing.}
        \label{fig:standardcnn}
    \end{subfigure}
    \hfill
    \begin{subfigure}[t]{0.48\columnwidth}
        \centering
        \includegraphics[width=\linewidth]{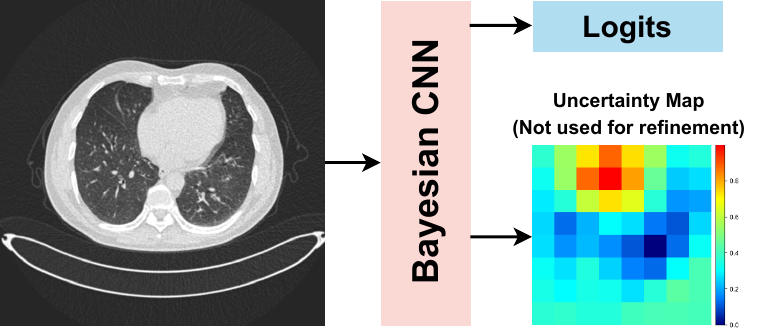}
        \caption{Bayesian CNN: Produces uncertainty maps but without focused refinement.}
        \label{fig:bayesiancnn}
    \end{subfigure}

    \vspace{0.5em}
    \begin{subfigure}[t]{0.98\columnwidth}
        \centering
        \includegraphics[width=\linewidth]{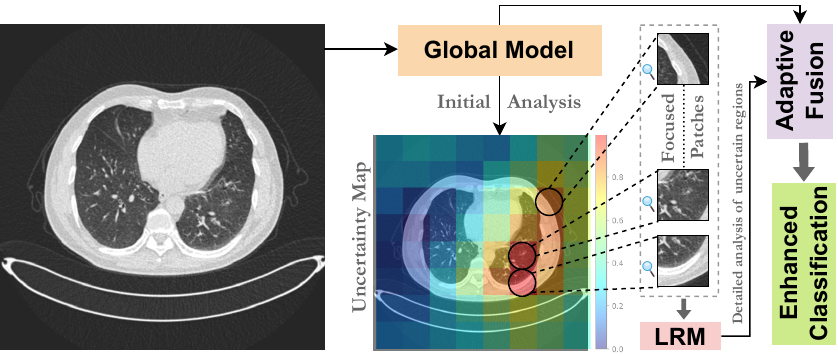}
        \caption{Uncertainty-Guided Progressive Learning (UGPL): Focuses computational resources on uncertain regions for enhanced classification.}
        \label{fig:ugpl}
    \end{subfigure}

    \caption{Comparison of medical image classification methods: unlike standard CNNs (\ref{fig:standardcnn}) and Bayesian CNNs (\ref{fig:bayesiancnn}) that process images uniformly, our proposed UGPL framework (\ref{fig:ugpl}) adaptively focuses on high-uncertainty regions for refined local analysis, combining global and local predictions via adaptive fusion.}
    \label{fig:teaser}
\end{figure}

\begin{figure*}[t]
    \centering
    \includegraphics[width=\textwidth]{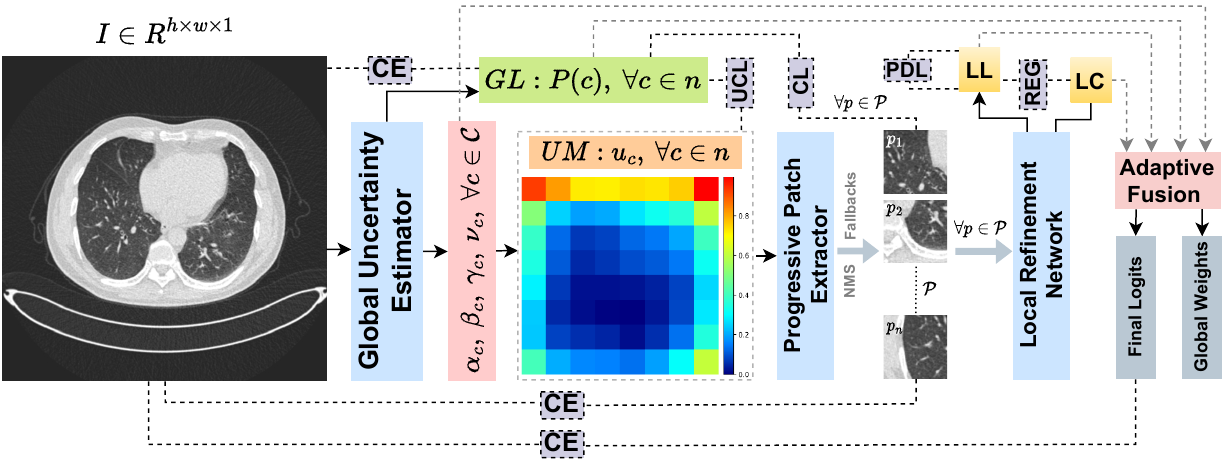}
    \caption{The UGPL architecture pipeline. Our framework processes an input CT image through a global uncertainty estimator to produce classification probabilities and an uncertainty map (left). The progressive patch extractor selects high-uncertainty regions for detailed analysis (center). These patches are processed by a local refinement network and combined with global predictions through an adaptive fusion module (right). Multiple loss functions (CE, UCC, CL, PDL, REG) are jointly optimized to ensure effective training of all components.}
    \label{fig:architecture}
\end{figure*}

Conventional CNNs apply identical convolutions across the image, ignoring regional diagnostic value. This is a limiting factor in medical imaging, where abnormalities may occupy a small fraction of the image. For instance, lung nodules or renal cysts often appear in confined regions that can be missed under uniform processing. Increasing model capacity or resolution globally is a possible workaround, but it incurs significant computational costs and delays in inference, both critical concerns in clinical practice. Moreover, such methods do not adapt their assessment based on uncertainty, unlike real practice, where analysis is refined based on perceived ambiguity. This gap in spatial adaptivity limits current models from capturing diagnostically important features in complex cases.

Several approaches have attempted to address aspects of this problem. Attention mechanisms~\cite{att1, att2, att3} and Region-based CNNs~\cite{rcn1,rcn2,rcn3} enable models to focus on specific image regions, but they typically identify regions based on learned patterns rather than diagnostic uncertainty. Bayesian neural networks~\cite{bayesiancnn,bayesiancnn2,bayesiancnn4} and Monte Carlo dropout techniques~\cite{mcd,mcd2,mcd3} offer uncertainty quantification in medical image analysis, producing pixel-wise uncertainty maps that highlight ambiguous regions. However, these methods primarily use uncertainty for confidence estimation or out-of-distribution detection, but not as analysis feedback. Progressive approaches in computer vision~\cite{pl1,pl2,pl3} process images in multiple stages of increasing resolution, but follow predetermined schedules rather than adapting based on detected uncertainty. While these methods offer partial solutions, they fail to integrate such uncertainty estimations with subsequent analysis refinement as done in practice.

In this paper, we introduce Uncertainty-Guided Progressive Learning (UGPL), a novel framework that mimics diagnostic behavior by performing global analysis followed by focused examination of uncertain regions (Figure~\ref{fig:teaser}). UGPL addresses limitations of uniform processing by dynamically allocating computational resources where needed. Our framework first employs a global uncertainty estimator to perform initial classification and generate pixel-wise uncertainty maps, then selects high-uncertainty regions for detailed analysis through a local refinement network. These multi-resolution analyses are combined via an adaptive fusion module that weights predictions based on confidence. Unlike existing methods that treat uncertainty merely as an output signal, UGPL explicitly uses it to guide computational focus, maintaining efficiency while improving performance on diagnostically challenging regions.

As shown in Figure~\ref{fig:architecture}, UGPL processes the input CT image to produce both classification probabilities and an uncertainty map that guides the extraction of high-uncertainty patches using non-maximum suppression. Each patch undergoes high-resolution analysis through a local refinement network, producing patch-specific classification scores and confidence estimates. The adaptive fusion module then integrates global and local predictions using learned weights based on their estimated reliability. Multiple specialized loss functions are jointly optimized, guiding components to work in tandem, adapt according to diagnostic difficulty, and improve performance over uniform processing.

To summarize, our main contributions are:

\begin{itemize}
\item a novel uncertainty-guided progressive learning framework that dynamically allocates computational resources to regions of high diagnostic ambiguity,
\item an evidential deep learning (EDL) approach that provides principled uncertainty quantification through Dirichlet distributions,
\item an adaptive patch extraction mechanism with non-maximum suppression that selects diverse, non-overlapping regions for detailed analysis, and
\item a multi-component loss formulation that jointly optimizes classification accuracy, uncertainty calibration, and spatial diversity.
\end{itemize}

\section{Related Works}
\label{sec:relatedWorks}

\textbf{Evidential Deep Learning in Medical Imaging.} EDL~\cite{edl} has been applied to various medical imaging tasks to model uncertainty and improve reliability. Early work integrated Dempster-Shafer Theory~\cite{dempster} into encoder-decoder architectures for 3D lymphoma segmentation, using voxel-level belief functions to improve accuracy and calibration over standard UNets~\cite{aref1}. In radiotherapy dose prediction, EDL showed that epistemic uncertainty correlates with prediction error, enabling confidence interval estimation for dose-volume histograms~\cite{aref3}. Extensions include region-based EDL with Dirichlet modeling for brain tumor delineation~\cite{aref4} and multi-view fusion architectures combining foundation models with uncertainty-aware layers to handle boundary ambiguity~\cite{aref5}.

Methods like EVIL~\cite{aref6, aref14} introduced efficient semi-supervised segmentation via uncertainty-guided consistency training, filtering unreliable pseudo-labels, and achieved strong results on ACDC and MM-WHS datasets. Multimodal and semi-supervised EDL variants further enhanced reliability. Dual-level evidential networks~\cite{aref11} and contextual discounting strategies~\cite{aref2,aref10} model modality trust in PET-CT and MRI fusion, improving voxel-level interpretability in tumor segmentation. Tri-branch frameworks like ETC-Net integrate evidential guidance with co-training to stabilize pseudo-labels in low-annotation regimes~\cite{aref12}. Beyond segmentation, EDL has been applied to classification, including three-way decision-making with EviDCNN~\cite{aref7} and out-of-distribution detection using evidential reconcile blocks~\cite{aref8}, demonstrating its versatility in uncertainty-aware diagnostics.

\textbf{Uncertainty Quantification in Medical Image Analysis.} Uncertainty quantification is widely used in medical image analysis to enhance model reliability amid noisy inputs, ambiguous boundaries, and limited annotations. Multi-decoder U-Net architectures capture inter-expert variability and generate uncertainty-calibrated segmentations~\cite{bref1}, while probabilistic U-Nets model aleatoric and epistemic uncertainties from annotation variability~\cite{bref6}. For classification, Bayesian deep learning models~\cite{bayesiancnn} like UA-ConvNet and BARF use Monte Carlo dropout~\cite{mcd} to estimate predictive uncertainty, achieving strong COVID-19 detection from chest X-rays~\cite{bref15,bref14}. Transfer learning quantifies epistemic uncertainty across modalities, detecting shifts between CT and X-rays~\cite{bref7}. Uncertainty-aware attention in hierarchical fusion networks, as in Hercules, improves performance across OCT, lung CT, and chest X-rays~\cite{bref10}.

In reconstruction, Bayesian deep unrolling jointly models image formation and uncertainty for MRI and CT~\cite{bref5}. Multimodal regression models like MoNIG estimate modality-specific uncertainties via Normal-Inverse Gamma mixtures for adaptive trust calibration~\cite{bref4}. For high-risk tasks such as COVID-19 classification, RCoNet combines mutual information maximization with ensemble dropout for robustness under distributional noise~\cite{bref3}. Distance-based out-of-distribution detection helps identify unreliable lung lesion segmentations~\cite{bref9}. Joint prediction confidence estimation aids data filtering and performance improvements in chest radiograph interpretation and ultrasound view classification~\cite{bref13}, underscoring the role of uncertainty quantification in reliable medical AI.

\section{Method}
\label{sec:method}

\begin{figure*}[t]
    \centering
    \includegraphics[width=\textwidth]{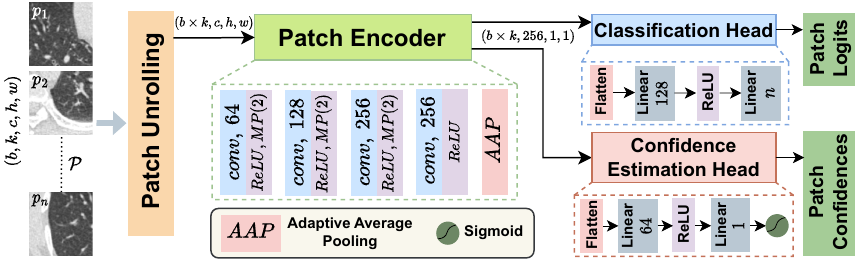}
    \caption{Architecture of the Local Refinement Network. The network processes extracted patches $(p_1, p_2, ..., p_n)$ through a patch encoder comprising four convolutional blocks with increasing feature dimensions (64→128→256→256) and adaptive average pooling. The encoded features are then fed into two parallel heads: a classification head that produces patch-specific logits, and a confidence estimation head that generates confidence scores used for weighted fusion of patch predictions.}
    \label{fig:local_refinement}
\end{figure*}

\subsection{Background}
\label{subsec:background}

Medical image classification requires both global contextual understanding and detailed examination of localized abnormalities. In this section, we establish the mathematical foundations for our uncertainty-guided approach.

\textbf{Evidential Deep Learning.} Traditional deep learning classifiers output class probabilities $p(y|\mathbf{x})$ directly but lack principled uncertainty quantification. Evidential Deep Learning (EDL) \cite{edl} addresses this by modeling a distribution over probabilities. For a classification problem with $C$ classes, EDL parameterizes a Dirichlet distribution $\text{Dir}(\mathbf{p}|\boldsymbol{\alpha})$ over the probability simplex, where $\boldsymbol{\alpha} = [\alpha_1, \alpha_2, ..., \alpha_C]$ are concentration parameters:

\begin{equation}
\text{Dir}(\mathbf{p}|\boldsymbol{\alpha}) = \frac{1}{B(\boldsymbol{\alpha})}\prod_{i=1}^{C} p_i^{\alpha_i-1}
\end{equation}

Here, $B(\boldsymbol{\alpha})$ is the multivariate beta function. The concentration parameters $\boldsymbol{\alpha}$ can be interpreted as evidence for each class, with $\alpha_i = e_i + 1$ where $e_i \geq 0$ represents the evidence for class $i$. The expected probability for class $i$ is given by $\mathbb{E}[p_i] = \frac{\alpha_i}{S}$, where $S = \sum_{i=1}^{C} \alpha_i$ is the Dirichlet strength.

EDL enables the quantification of two types of uncertainty: \textit{aleatoric uncertainty} (data uncertainty) and \textit{epistemic uncertainty} (model uncertainty). For a Dirichlet distribution, the total predictive uncertainty can be computed as:

\begin{equation}
\mathcal{U}_{\text{total}} = \sum_{i=1}^{C} \frac{\alpha_i}{S} \left(1 - \frac{\alpha_i}{S}\right) \frac{1}{S+1}
\end{equation}

This captures both the entropy of the expected categorical distribution (first term) and the additional uncertainty from the Dirichlet distribution itself (second term).

\subsection{Global Uncertainty Estimation and Evidential Learning}
\label{subsec:global_uncertainty}

Our global uncertainty estimator produces initial class predictions and generates a spatial uncertainty map to guide patch selection through evidential learning.

\textbf{Global Model Architecture.}
Given an input CT image $\mathbf{I} \in \mathbb{R}^{H \times W \times 1}$, we employ a ResNet backbone~\cite{resnet} $\mathcal{F}_\theta$ to extract feature maps $\mathbf{F} \in \mathbb{R}^{h \times w \times d}$. To accommodate grayscale CT images, we modify the first convolutional layer to accept single-channel inputs while preserving pretrained weights by averaging across RGB channels. The feature maps are processed by two parallel heads: a classification head $\mathcal{C}_\phi$ and an evidence head $\mathcal{E}_\psi$. The classification head applies global average pooling followed by a fully connected layer to produce class logits: $\mathbf{z}_g = \mathcal{C}_\phi(\mathbf{F})$.

\textbf{Evidential Uncertainty Estimation.}
The evidence head $\mathcal{E}_\psi$ generates pixel-wise Dirichlet concentration parameters that quantify uncertainty at each spatial location: $\mathbf{E} = \mathcal{E}_\psi(\mathbf{F}) \in \mathbb{R}^{h \times w \times 4C}$, where $\mathbf{E}$ encodes four parameters $(\alpha, \beta, \gamma, \nu)$ for each class at each location. Following subjective logic principles~\cite{josang2016subjective}, we parameterize the Dirichlet distribution as:

\begin{equation}
\alpha_{i,j,c} = \beta_{i,j,c} \cdot \nu_{i,j,c} + 1
\end{equation}

where $\beta_{i,j,c}$ represents the inverse of uncertainty, $\nu_{i,j,c}$ represents the mass belief, and we constrain $\sum_{c=1}^{C} \nu_{i,j,c} = 1$. From these parameters, we compute the pixel-wise uncertainty map $\mathbf{U} \in \mathbb{R}^{h \times w}$ by aggregating uncertainty across all classes:

\begin{equation}
\mathbf{U}_{i,j} = \frac{1}{C} \sum_{c=1}^{C} \left(\frac{1}{\alpha_{i,j,c}} + \frac{\beta_{i,j,c}}{\alpha_{i,j,c}(\alpha_{i,j,c}+1)}\right)
\end{equation}

The first term accounts for aleatoric uncertainty, and the second for epistemic uncertainty. We normalize the uncertainty map to $[0,1]$ for easier interpretation and subsequent processing.

\subsection{Uncertainty-Guided Patch Selection and Local Refinement}
\label{subsec:patch_selection}

\textbf{Progressive Patch Extraction.}
Given an input image $\mathbf{I} \in \mathbb{R}^{H \times W \times 1}$ and its corresponding uncertainty map $\hat{\mathbf{U}} \in \mathbb{R}^{h \times w}$, we first upsample the uncertainty map to match the input resolution: $\mathbf{U}' = \mathcal{U}(\hat{\mathbf{U}}, (H, W))$. Our objective is to extract $K$ patches of size $P \times P$ from the input image based on an uncertainty-guided selection process. We formulate this as a greedy algorithm that selects patches from the original image at locations corresponding to maxima in the uncertainty map $\mathbf{U}'$. The first patch is centered at the global maximum uncertainty:
\begin{align}
(x_1, y_1) = \arg\max_{(x,y)} \mathbf{U}'_{x:x+P, y:y+P}
\end{align}
For subsequent patches, we introduce a spatial penalty term that encourages diversity by maintaining minimum distance from previously selected locations:
\begin{align}
(x_k, y_k) = \arg\max_{(x,y)} \Big[\, 
    & \mathbf{U}'_{x:x+P,\, y:y+P} \notag \\
    & - \lambda \cdot \min_{i<k} d((x,y),\, (x_i, y_i)) 
\,\Big]
\end{align}
This sequential optimization ensures that each new patch maximizes uncertainty while preventing redundant selection of nearby regions. We implement this efficiently using a non-maximum suppression approach, applying a Gaussian suppression kernel after selecting each patch. Algorithm~\ref{alg:patch_extraction} details our complete patch extraction procedure, including practical considerations for edge cases.

\begin{algorithm}[t]
\caption{Uncertainty-Guided Patch Extraction}
\label{alg:patch_extraction}
\begin{algorithmic}[1]
\REQUIRE Input image $\mathbf{I} \in \mathbb{R}^{H \times W \times 1}$, Uncertainty map $\hat{\mathbf{U}} \in \mathbb{R}^{h \times w}$, Patch size $P$, Number of patches $K$
\ENSURE Set of patches $\{\mathbf{P}_1, \mathbf{P}_2, \ldots, \mathbf{P}_K\}$, Patch coordinates $\{(x_1, y_1), (x_2, y_2), \ldots, (x_K, y_K)\}$

\STATE $\mathbf{U}' \leftarrow \mathcal{U}(\hat{\mathbf{U}}, (H, W))$ \COMMENT{Upsample uncertainty map}
\STATE Initialize patch coordinates list $\mathcal{C} \leftarrow \{\}$
\STATE $\mathbf{M} \leftarrow$ zeros($H, W$) \COMMENT{Mask for selected regions}

\FOR{$k = 1$ to $K$}
   \STATE $\mathbf{V} \leftarrow \mathbf{U}' \odot (1 - \mathbf{M})$ \COMMENT{Apply mask to uncertainty map}
   
   \IF{$\max(\mathbf{V}) > 0$}
       \STATE $(y_k, x_k) \leftarrow \arg\max_{(y,x)} \mathbf{V}$ \COMMENT{Find maximum uncertainty location}
   \ELSE
       \STATE $(y_k, x_k) \leftarrow$ random valid location \COMMENT{Fallback: random selection}
   \ENDIF
   
   \STATE $x_k \leftarrow \max(0, \min(x_k, W-P))$ \COMMENT{Ensure patch fits within image}
   \STATE $y_k \leftarrow \max(0, \min(y_k, H-P))$
   
   \STATE $\mathcal{C} \leftarrow \mathcal{C} \cup \{(x_k, y_k)\}$ \COMMENT{Add coordinates to list}
   
   \STATE $\mathbf{M}_{y_k-M:y_k+P+M, x_k-M:x_k+P+M} \leftarrow 1$ \COMMENT{Update mask with margin $M$}
   
   \STATE $\mathbf{P}_k \leftarrow \mathbf{I}_{y_k:y_k+P, x_k:x_k+P}$ \COMMENT{Extract patch}
   
   \IF{$\mathbf{P}_k$ size $\neq (P, P)$}
       \STATE $\mathbf{P}_k \leftarrow$ Resize($\mathbf{P}_k$, $(P, P)$) \COMMENT{Ensure consistent size}
   \ENDIF
\ENDFOR

\RETURN $\{\mathbf{P}_1, \mathbf{P}_2, \ldots, \mathbf{P}_K\}$, $\mathcal{C}$
\end{algorithmic}
\end{algorithm}

\textbf{Local Refinement Network.}After extracting $K$ patches, we process each independently using a local refinement network with three components (Figure~\ref{fig:local_refinement}): a feature extractor, a classification head, and a confidence estimation head. The feature extractor $\mathcal{L}_f$ processes each patch to obtain local feature vectors: $\mathbf{f}_k = \mathcal{L}_f(\mathbf{P}_k) \in \mathbb{R}^{d_l}$. The classification head maps these features to class logits: $\mathbf{z}_{l,k} = \mathcal{L}_c(\mathbf{f}_k) \in \mathbb{R}^C$, while the confidence estimation head produces a scalar confidence score: $c_k = \mathcal{L}_\text{conf}(\mathbf{f}_k) \in [0, 1]$.

The confidence score allows the model to express uncertainty about individual patch predictions and weights their contribution in the final classification. The combined local prediction is computed as a confidence-weighted average: $\mathbf{z}_l = \frac{\sum_{k=1}^K c_k \cdot \mathbf{z}_{l,k}}{\sum_{k=1}^K c_k + \epsilon}$, where $\epsilon$ is a small constant for numerical stability. This naturally reduces the contribution of low-confidence patches, functioning as an implicit attention mechanism that focuses on the most discriminative regions.

\subsection{Adaptive Fusion and Training Objectives}
\label{subsec:fusion_training}

\textbf{Adaptive Fusion Module.}
Given the global logits $\mathbf{z}_g \in \mathbb{R}^C$ and uncertainty map $\hat{\mathbf{U}} \in \mathbb{R}^{h \times w}$ from the global model, and local logits $\mathbf{z}_l \in \mathbb{R}^C$ with patch confidence scores $\{c_1, c_2, \ldots, c_K\}$ from the local refinement network, our adaptive fusion module dynamically balances global and local predictions.

We compute a scalar global uncertainty $u_g = \frac{1}{h \cdot w} \sum_{i=1}^h \sum_{j=1}^w \hat{\mathbf{U}}_{i,j}$ to quantify the overall confidence of the global model. The fusion network $\mathcal{F}_\omega$ takes as input $[\mathbf{z}_g, u_g]$ and outputs a fusion weight $w_g = \mathcal{F}_\omega([\mathbf{z}_g, u_g])$, implemented as a multi-layer perceptron with sigmoid activation. The fused logits are computed as $\mathbf{z}_f = w_g \cdot \mathbf{z}_g + (1 - w_g) \cdot \mathbf{z}_l$. This adaptive weighting relies more on global features when the global model is confident, and more on local features when uncertainty is high.

\begin{table*}[t]
\centering
\caption{Comparison of our UGPL approach with state-of-the-art classification models across three CT datasets. Results on the COVID dataset for CRNet~\cite{coviddataset} are as reported in the paper. Best results are in \textbf{\textcolor{red}{red}}, second-best in \textbf{\textcolor{blue}{blue}}, and third-best in \textbf{\textcolor{green}{green}}.}
\label{tab:sota_comparison}
\resizebox{\textwidth}{!}{
\setlength{\tabcolsep}{8pt}
\begin{tabular}{l|cc|cc|cc}
\hline
\multirow{2}{*}{Models} & \multicolumn{2}{c|}{Kidney Abnormalities} & \multicolumn{2}{c|}{Lung Cancer Type} & \multicolumn{2}{c}{COVID Presence} \\
\cline{2-7}
 & Accuracy & F1 & Accuracy & F1 & Accuracy & F1 \\
\hline
ShuffleNetV2~\cite{shufflenet} & 0.96\,$\pm$\,0.0085 & 0.95\,$\pm$\,0.0092 & 0.94\,$\pm$\,0.0127 & 0.91\,$\pm$\,0.0143 & 0.69\,$\pm$\,0.0234 & 0.67\,$\pm$\,0.0251 \\
VGG16~\cite{vgg} & 0.89\,$\pm$\,0.0156 & 0.88\,$\pm$\,0.0173 & \textcolor{blue}{\textbf{0.95\,$\pm$\,0.0098}} & 0.91\,$\pm$\,0.0165 & 0.48\,$\pm$\,0.0287 & 0.47\,$\pm$\,0.0306 \\
ConvNeXt~\cite{convnext} & 0.81\,$\pm$\,0.0189 & 0.80\,$\pm$\,0.0195 & \textcolor{green}{\textbf{0.95\,$\pm$\,0.0076}} & \textcolor{green}{\textbf{0.95\,$\pm$\,0.0084}} & 0.61\,$\pm$\,0.0267 & 0.59\,$\pm$\,0.0278 \\
DenseNet121~\cite{densenet} & 0.94\,$\pm$\,0.0102 & 0.93\,$\pm$\,0.0118 & 0.90\,$\pm$\,0.0171 & 0.89\,$\pm$\,0.0176 & \textcolor{blue}{\textbf{0.78\,$\pm$\,0.0198}} & \textcolor{blue}{\textbf{0.76\,$\pm$\,0.0213}} \\
DenseNet201~\cite{densenet} & 0.95\,$\pm$\,0.0093 & 0.94\,$\pm$\,0.0106 & 0.84\,$\pm$\,0.0203 & 0.83\,$\pm$\,0.0218 & \textcolor{green}{\textbf{0.76\,$\pm$\,0.0206}} & 0.74\,$\pm$\,0.0229 \\
EfficientNetB0~\cite{effnet} & 0.95\,$\pm$\,0.0078 & 0.94\,$\pm$\,0.0089 & 0.95\,$\pm$\,0.0081 & \textcolor{blue}{\textbf{0.95\,$\pm$\,0.0073}} & 0.73\,$\pm$\,0.0221 & 0.71\,$\pm$\,0.0238 \\
MobileNetV2~\cite{mobilenet} & 0.87\,$\pm$\,0.0179 & 0.85\,$\pm$\,0.0195 & 0.70\,$\pm$\,0.0267 & 0.69\,$\pm$\,0.0283 & 0.70\,$\pm$\,0.0241 & 0.68\,$\pm$\,0.0256 \\
ViT~\cite{visiontransformer} & 0.94\,$\pm$\,0.0154 & 0.92\,$\pm$\,0.0167 & 0.51\,$\pm$\,0.0389 & 0.22\,$\pm$\,0.0456 & 0.56\,$\pm$\,0.0312 & 0.55\,$\pm$\,0.0318 \\
Swin~\cite{swintransformer} & 0.68\,$\pm$\,0.0298 & 0.40\,$\pm$\,0.0421 & 0.60\,$\pm$\,0.0334 & 0.41\,$\pm$\,0.0398 & 0.53\,$\pm$\,0.0331 & 0.53\,$\pm$\,0.0329 \\
DeiT~\cite{deit} & 0.92\,$\pm$\,0.0162 & 0.90\,$\pm$\,0.0178 & 0.66\,$\pm$\,0.0312 & 0.46\,$\pm$\,0.0387 & 0.44\,$\pm$\,0.0356 & 0.35\,$\pm$\,0.0412 \\
CoaT~\cite{coat} & \textcolor{blue}{\textbf{0.98\,$\pm$\,0.0067}} & \textcolor{blue}{\textbf{0.98\,$\pm$\,0.0072}} & 0.95\,$\pm$\,0.0089 & 0.93\,$\pm$\,0.0112 & 0.68\,$\pm$\,0.0254 & 0.66\,$\pm$\,0.0267 \\
CrossViT~\cite{crossvit} & \textcolor{green}{\textbf{0.97\,$\pm$\,0.0087}} & \textcolor{green}{\textbf{0.97\,$\pm$\,0.0094}} & 0.58\,$\pm$\,0.0356 & 0.39\,$\pm$\,0.0423 & 0.62\,$\pm$\,0.0289 & 0.48\,$\pm$\,0.0378 \\
CRNet~\cite{coviddataset} & - & - & - & - & 0.73\,$\pm$\,0.0218 & \textcolor{green}{\textbf{0.76\,$\pm$\,0.0203}} \\
\rowcolor{gray!30}
UGPL (Ours) & \textcolor{red}{\textbf{0.99\,$\pm$\,0.0023}} & \textcolor{red}{\textbf{0.99\,$\pm$\,0.0031}} & \textcolor{red}{\textbf{0.98\,$\pm$\,0.0047}} & \textcolor{red}{\textbf{0.97\,$\pm$\,0.0052}} & \textcolor{red}{\textbf{0.81\,$\pm$\,0.0134}} & \textcolor{red}{\textbf{0.79\,$\pm$\,0.0147}} \\
\hline
\end{tabular}
}
\end{table*}

\textbf{Multi-component Loss Function.}
Our training uses a comprehensive loss function combining several objectives:

\begin{equation}
\begin{split}
\mathcal{L}_\text{total} =\ 
    & \lambda_f \mathcal{L}_\text{fused}
    + \lambda_g \mathcal{L}_\text{global} \\
    & + \lambda_l \mathcal{L}_\text{local}
    + \lambda_u \mathcal{L}_\text{uncertainty} \\
    & + \lambda_c \mathcal{L}_\text{consistency}
    + \lambda_{\text{conf}} \mathcal{L}_\text{confidence} \\
    & + \lambda_d \mathcal{L}_\text{diversity}
\end{split}
\end{equation}

\textbf{Classification Losses.} We apply cross-entropy loss to predictions from each component: $\mathcal{L}_\text{fused}$ for the fused predictions, $\mathcal{L}_\text{global}$ for global predictions, and $\mathcal{L}_\text{local}$ averaged across all patch predictions.

\textbf{Auxiliary Losses.} We also use several auxiliary components to ensure effective training: (1) $\mathcal{L}_\text{uncertainty}$ calibrates the uncertainty map to reflect prediction errors; (2) $\mathcal{L}_\text{consistency}$ promotes agreement between global and local predictions using KL divergence weighted by patch confidence; (3) $\mathcal{L}_\text{confidence}$ aligns patch confidence scores with prediction accuracy; and (4) $\mathcal{L}_\text{diversity}$ encourages diversity among patch predictions through cosine similarity penalization. 

\section{Experiments}
\label{sec:experiments}

\subsection{Experimental Setup}
\label{subsec:exp_setup}

\textbf{Datasets.} We conduct experiments on three CT image datasets: the kidney disease diagnosis dataset~\cite{kidneydataset} (multiclass: normal, cyst, tumor, stone), the IQ-OTH/NCCD lung cancer dataset~\cite{lungdataset1,lungdataset2,lungdataset3} (multiclass: benign, malignant, normal), and the UCSD-AI4H COVID-CT dataset~\cite{coviddataset} (binary: COVID, non-COVID). All images are resized to $256 \times 256$ resolution during preprocessing and normalized using the respective dataset's mean and standard deviation.

\begin{table}[t]
\centering
\caption{Analysis of individual component performance in our UGPL framework across the three datasets. The \colorbox{gray!30}{shaded row} corresponds to our baseline configuration.}
\label{tab:component_analysis}
\resizebox{\columnwidth}{!}{
\begin{tabular}{l|cc|cc|cc}
\hline
\multirow{2}{*}{Model Type} & \multicolumn{2}{c|}{COVID Presence} & \multicolumn{2}{c|}{Lung Cancer Type} & \multicolumn{2}{c}{Kidney Abnormalities} \\
\cline{2-7}
 & Accuracy & F1 & Accuracy & F1 & Accuracy & F1 \\
\hline
Global Model & 0.7108 & 0.7078 & 0.9617 & 0.9611 & 0.9811 & 0.9746 \\
Local Model & 0.6486 & 0.6343 & 0.5122 & 0.2258 & 0.4057 & 0.1443 \\
\rowcolor{gray!30}
Fused Model & 0.8108 & 0.7903 & 0.9817 & 0.9764 & 0.9971 & 0.9946 \\
\hline
\end{tabular}
}
\end{table}

\begin{figure}[t]
\centering
\includegraphics[width=\linewidth]{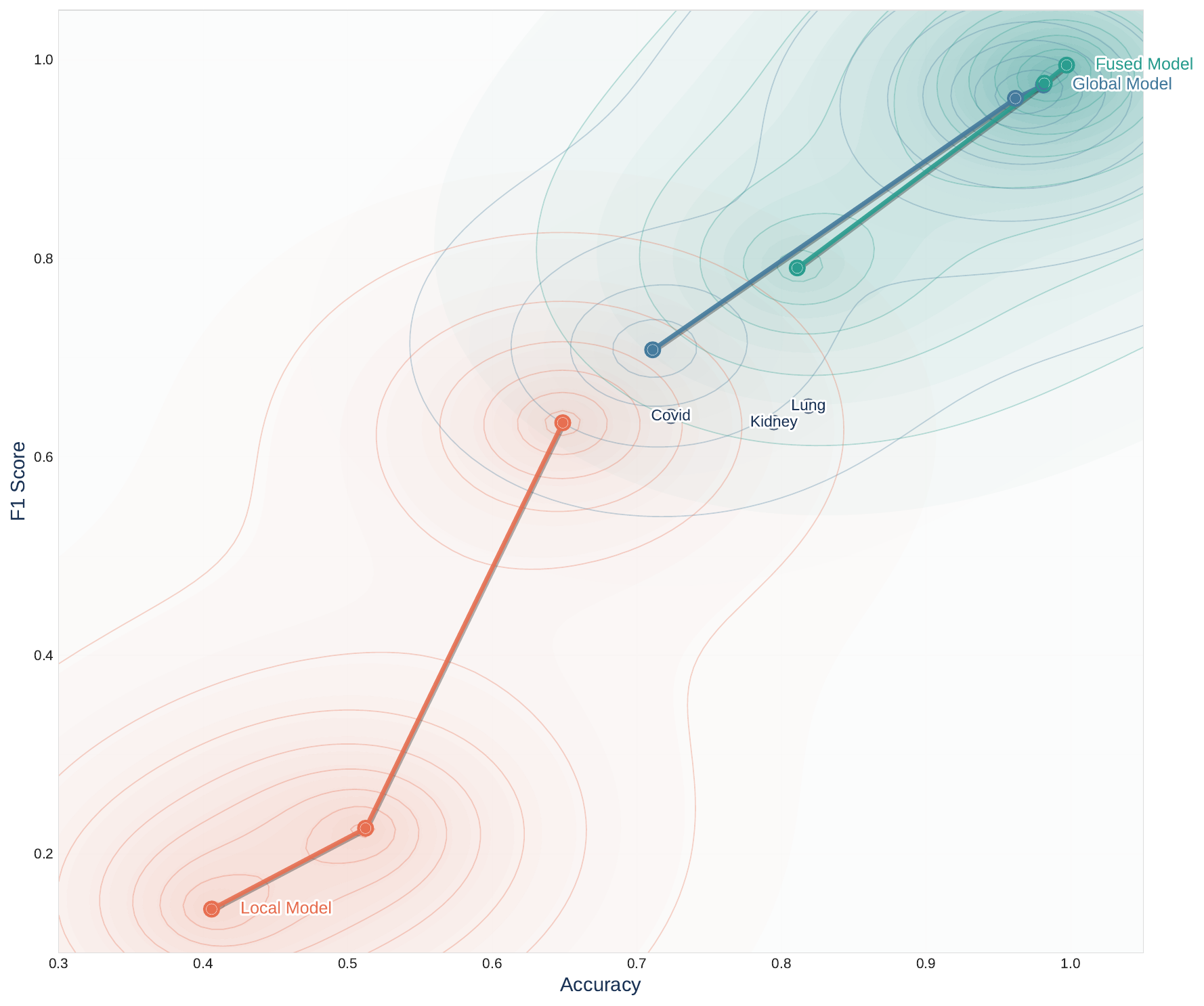}
\caption{Performance trends of model components across datasets. Accuracy (x-axis) and F1 score (y-axis) define trajectories from LM to GM to FM, with contour lines indicating performance density.}
\label{fig:component_flow}
\end{figure}

\textbf{Implementation Details.} All models were trained for 100 epochs using Adam optimizer~\cite{AdamOpt} with learning rate $1 \times 10^{-4}$, weight decay $1 \times 10^{-4}$, batch size 96, and cosine decay scheduling~\cite{scheduler}. Standard augmentations included flips, rotations, affine transformations, and contrast adjustments. Dataset-specific ResNet~\cite{resnet} backbones were used with varying patch configurations. The multi-component loss function employed weighted components for fused (1.0), global/local (0.5), uncertainty (0.3), consistency (0.2), and confidence/diversity losses (0.1). 

\begin{figure*}[t]
\centering
\includegraphics[width=\linewidth]{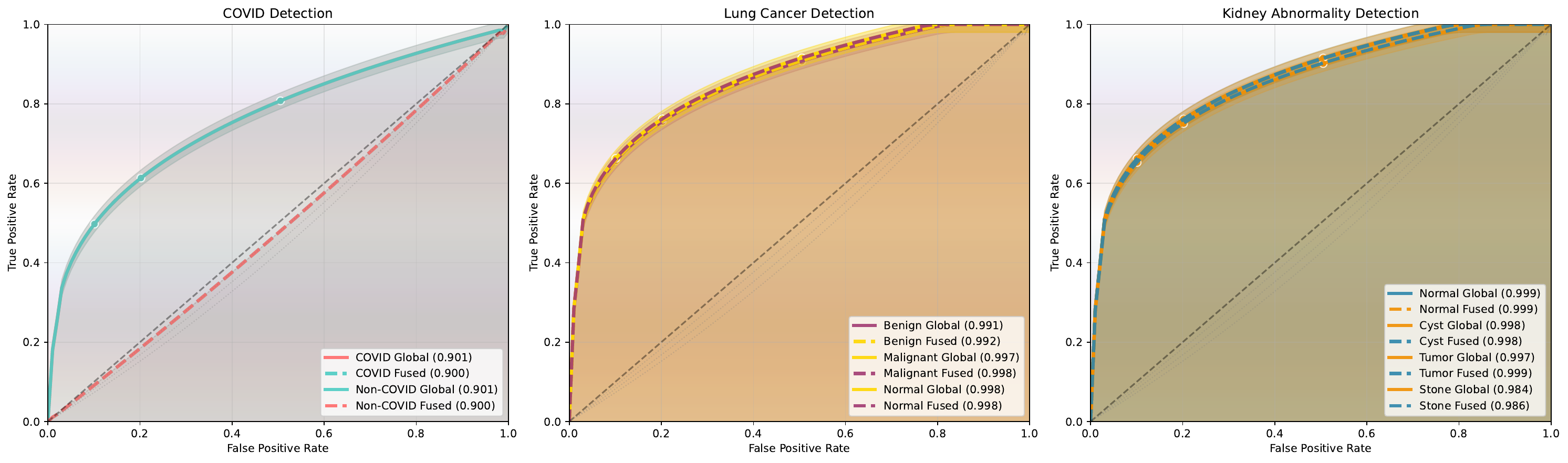}
\caption{ROC curves comparing global and fused model performance across datasets. The FM consistently maintains or improves the already high AUC values of the GM across all classes and datasets.}
\label{fig:roc_curves}
\end{figure*}

\subsection{Performance Evaluation}
\label{subsec:performance_evaluation}

Table~\ref{tab:sota_comparison} shows the performance of our method against a range of CNN and transformer-based models. These include lightweight CNNs (MobileNetV2~\cite{mobilenet}, ShuffleNetV2~\cite{shufflenet}), standard convolutional baselines (VGG16~\cite{vgg}, DenseNet121/201~\cite{densenet}, EfficientNetB0~\cite{effnet}, ConvNeXt~\cite{convnext}), and recent transformer-based architectures (ViT~\cite{visiontransformer}, Swin~\cite{swintransformer}, DeiT~\cite{deit}, CoaT~\cite{coat}, CrossViT~\cite{crossvit}). We compare our UGPL approach across three CT classification tasks. For all models, we report accuracy, macro-averaged F1 score, and include ROC-AUC visualizations for further analysis.

On the kidney abnormality dataset~\cite{kidneydataset}, UGPL achieves the highest accuracy and F1-score at 99\% (±0.0023, ±0.0031). Among CNNs, CoaT~\cite{coat}, CrossViT~\cite{crossvit}, and EfficientNetB0~\cite{effnet} follow with F1 between 94--98\%, all with low variance. MobileNetV2~\cite{mobilenet} and VGG16~\cite{vgg} fall below 89\%. Transformer models like ViT~\cite{visiontransformer} and Swin~\cite{swintransformer} show lower F1 and higher deviations, with Swin dropping to 40\% F1 (±0.0421).

On the IQ-OTH/NCCD dataset~\cite{lungdataset1,lungdataset2,lungdataset3}, UGPL reports 97\% F1 (±0.0052), the highest overall. CNNs such as EfficientNetB0~\cite{effnet} and ConvNeXt~\cite{convnext} reach 95\% F1 with low variance. CoaT~\cite{coat} and VGG16~\cite{vgg} follow closely, while transformer models like DeiT~\cite{deit} and Swin~\cite{swintransformer} perform poorly, with F1 below 50\% and higher spread. Variance is generally higher for transformers, with less consistent learning across folds.

For COVID classification, UGPL leads with 79\% F1 (±0.0147), followed by DenseNet121~\cite{densenet} and CRNet~\cite{coviddataset} at 76\%. EfficientNetB0~\cite{effnet} and DenseNet201~\cite{densenet} also perform in the 71--74\% range. Most transformer-based models, including ViT~\cite{visiontransformer}, Swin~\cite{swintransformer}, and DeiT~\cite{deit}, remain under 60\% F1 with variances exceeding ±0.03. These models also show less consistency across folds, with notably higher performance fluctuations.

\subsection{Component Analysis}
\label{subsec:component_analysis}

\begin{table}[t]
\centering
\caption{Ablation study of different model component configurations across the three datasets. The \colorbox{gray!30}{shaded row} corresponds to our baseline configuration.}
\label{tab:ablation_configs}
\resizebox{\columnwidth}{!}{
\begin{tabular}{l|cc|cc|cc}
\hline
\multirow{2}{*}{Configuration} & \multicolumn{2}{c|}{COVID Presence} & \multicolumn{2}{c|}{Lung Cancer Type} & \multicolumn{2}{c}{Kidney Abnormalities} \\
\cline{2-7}
 & Accuracy & F1 & Accuracy & F1 & Accuracy & F1 \\
\hline
Global-only & 0.2535 & 0.1495 & 0.5000 & 0.3890 & 0.5676 & 0.5545 \\
No UG & 0.2363 & 0.1536 & 0.4634 & 0.3764 & 0.5766 & 0.5558 \\
Fixed Patches & 0.2347 & 0.1533 & 0.4573 & 0.3731 & 0.5766 & 0.5697 \\
\rowcolor{gray!30}
Full Model & 0.8108 & 0.7903 & 0.9817 & 0.9764 & 0.9971 & 0.9945 \\
\hline
\end{tabular}
}
\end{table}

Table~\ref{tab:component_analysis} shows the contribution of each component in our UGPL framework. The global model (GM), performing whole-image classification, achieves strong performance on the Kidney and Lung datasets (98.11\% and 96.17\% accuracy). The local model (LM), using only patch-based classification, shows significantly lower performance when used independently. The fused model (FM), integrating both predictions through our adaptive fusion mechanism, consistently outperforms individual components.

The performance gap between GM and FM is most evident in COVID-19 detection, with FM reaching 81.08\% accuracy compared to 71.08\% for GM. This reflects the benefit of incorporating localized analysis in tasks where global patterns are less prominent. For kidney abnormality detection, FM also improves over GM (99.71\% vs. 98.11\%), showing that local refinement can still enhance outcomes even when global features are already effective.

\begin{table*}[t]
\centering
\caption{Performance comparison of different loss weight configurations across datasets. Loss component weights: Fused ($\lambda_f$), Global ($\lambda_g$), Local ($\lambda_l$), Uncertainty ($\lambda_u$), Consistency ($\lambda_c$), Confidence ($\lambda_{\text{co}}$), Diversity ($\lambda_d$). \colorbox{gray!30}{Configuration C1} represents our baseline model with balanced weights. Best results are in \textcolor{red}{\textbf{red}}, second-best in \textcolor{blue}{\textbf{blue}}, and third-best in \textcolor{green}{\textbf{green}}.}
\label{tab:loss_weights}
\resizebox{\textwidth}{!}{
\begin{tabular}{l|ccccccc|cc|cc|cc}
\hline
\multirow{2}{*}{Configuration} & \multicolumn{7}{c|}{Loss Weights} & \multicolumn{2}{c|}{COVID Presence} & \multicolumn{2}{c|}{Lung Cancer Type} & \multicolumn{2}{c}{Kidney Abnormalities} \\
\cline{2-14}
& $\lambda_f$ & $\lambda_g$ & $\lambda_l$ & $\lambda_u$ & $\lambda_c$ & $\lambda_{\text{co}}$ & $\lambda_d$ & Accuracy & F1 & Accuracy & F1 & Accuracy & F1 \\
\hline
\rowcolor{gray!30}
C1: Baseline & 1.0 & 0.5 & 0.5 & 0.3 & 0.2 & 0.1 & 0.1 & \textcolor{blue}{\textbf{0.8108}} & \textcolor{blue}{\textbf{0.7903}} & \textcolor{red}{\textbf{0.9817}} & \textcolor{red}{\textbf{0.9764}} & \textcolor{red}{\textbf{0.9971}} & \textcolor{red}{\textbf{0.9945}} \\
C2: Local Emphasis & 1.0 & 0.3 & 0.7 & 0.3 & 0.2 & 0.1 & 0.1 & 0.7946 & 0.7758 & 0.9695 & 0.9641 & 0.9928 & 0.9903 \\
C3: Global-Centric & 1.0 & 0.7 & 0.3 & 0.3 & 0.2 & 0.1 & 0.1 & 0.7568 & 0.7402 & 0.9634 & 0.9576 & 0.9876 & 0.9832 \\
C4: Uncertainty Focus & 1.0 & 0.5 & 0.5 & 0.6 & 0.2 & 0.1 & 0.1 & \textcolor{red}{\textbf{0.8243}} & \textcolor{red}{\textbf{0.8057}} & 0.9756 & 0.9687 & \textcolor{blue}{\textbf{0.9953}} & \textcolor{blue}{\textbf{0.9931}} \\
C5: Consistency-Driven & 1.0 & 0.5 & 0.5 & 0.3 & 0.5 & 0.1 & 0.1 & 0.7892 & 0.7689 & \textcolor{green}{\textbf{0.9786}} & \textcolor{green}{\textbf{0.9723}} & 0.9913 & 0.9889 \\
C6: Balanced High & 1.0 & 0.5 & 0.5 & 0.4 & 0.4 & 0.2 & 0.2 & \textcolor{green}{\textbf{0.8051}} & \textcolor{green}{\textbf{0.7836}} & \textcolor{blue}{\textbf{0.9801}} & \textcolor{blue}{\textbf{0.9739}} & \textcolor{green}{\textbf{0.9942}} & \textcolor{green}{\textbf{0.9918}} \\
C7: Diversity-Enhanced & 1.0 & 0.5 & 0.5 & 0.3 & 0.2 & 0.1 & 0.4 & 0.7784 & 0.7569 & 0.9667 & 0.9602 & 0.9895 & 0.9856 \\
C8: Confidence-Calibrated & 1.0 & 0.5 & 0.5 & 0.3 & 0.2 & 0.4 & 0.1 & 0.7973 & 0.7798 & 0.9753 & 0.9695 & 0.9923 & 0.9891 \\
C9: Conservative & 0.5 & 0.25 & 0.25 & 0.15 & 0.1 & 0.05 & 0.05 & 0.7486 & 0.7312 & 0.9581 & 0.9524 & 0.9837 & 0.9803 \\
C10: Aggressive & 2.0 & 1.0 & 1.0 & 0.6 & 0.4 & 0.2 & 0.2 & 0.8023 & 0.7827 & 0.9728 & 0.9674 & 0.9932 & 0.9907 \\
\hline
\end{tabular}
}
\end{table*}

The LM performs poorly across all tasks, particularly for kidney abnormalities (40.57\%) and lung cancer classification (51.22\%), as local patches alone lack sufficient context and focus on irrelevant regions without global guidance. Figure~\ref{fig:component_flow} shows performance trends across tasks, with COVID-19 detection showing the most significant gains from LM to FM. ROC curves (Figure~\ref{fig:roc_curves}) show that for COVID-19, GM and FM achieve similar AUC scores (0.901 vs. 0.900). For lung cancer, FM achieves slight improvements across classes, especially for benign cases (0.991 vs. 0.992). For kidney cases, FM improves performance for most classes, including kidney stones (0.984 vs. 0.986).

\subsection{Ablation Study}
\label{subsec:ablation}

We conduct extensive ablation studies to evaluate the impact of different components in our UGPL framework. We focus on three key aspects: 1) the contribution of each component in the progressive learning pipeline, 2) the influence of patch extraction parameters on model performance, and 3) the effect of varying loss term coefficients in our multi-component optimization objective. We retain the best-performing ResNet variant~\cite{resnet} from our initial evaluations for all experiments.

\subsubsection{Component Ablation}

To analyze the contribution of each component in our progressive learning framework, we compare four configurations: (1) a global-only setup that uses the global uncertainty estimator without local refinement; (2) a no uncertainty guidance (No UG) variant, where patches are selected randomly instead of using uncertainty maps; (3) a fixed patches configuration that uses predefined patch locations rather than adaptive selection; and (4) the full model, which includes all components of the UGPL framework.

Table~\ref{tab:ablation_configs} shows our full model consistently outperforming all reduced variants by substantial F1 margins. On the COVID dataset, all ablations cause dramatic performance drops, with the global-only variant achieving only 14.95\% F1. For lung cancer detection, the full model obtains 97.64\% F1, while the global-only setup drops to 34.19\%. The kidney dataset shows smaller yet significant gaps, with the full model reaching 99.6\% F1 versus 58.7\% for the best ablated configuration (fixed patches). Interestingly, No UG and fixed patches sometimes perform worse than the global-only model, showing that naively adding local components without proper guidance can be detrimental and highlighting the importance of uncertainty-guided patch selection.

\begin{table}[t]
\centering
\caption{F1 Scores across patch sizes and number of extracted patches. \textbf{Bolded values} indicate results from C1 configuration.}
\label{tab:patch_params}
\begin{tabular}{c|c|ccc}
\toprule
\textbf{Patch Size} & \textbf{Patches} & \textbf{Kidney} & \textbf{Lung} & \textbf{COVID} \\
\midrule
\multirow{3}{*}{32} 
  & 2 & 0.9586 & 0.8869 & 0.7161 \\
  & 3 & 0.9673 & 0.9195 & 0.7368 \\
  & 4 & 0.9541 & 0.8756 & 0.7454 \\
\midrule
\multirow{3}{*}{64} 
  & 2 & 0.9824 & \textbf{0.9764} & 0.7521 \\
  & 3 & \textbf{0.9945} & 0.8671 & 0.7368 \\
  & 4 & 0.9765 & 0.9343 & \textbf{0.7903} \\
\midrule
\multirow{3}{*}{96} 
  & 2 & 0.9622 & 0.8712 & 0.7372 \\
  & 3 & 0.9701 & 0.9099 & 0.7262 \\
  & 4 & 0.9418 & 0.8717 & 0.6505 \\
\bottomrule
\end{tabular}
\end{table}

\subsubsection{Loss Component Weights}

Table~\ref{tab:loss_weights} compares ten loss weight configurations across datasets. The baseline configuration (C1) with balanced weights performs best overall (fused: 1.0, global/local: 0.5 each, uncertainty: 0.3, consistency: 0.2, confidence/diversity: 0.1 each). Configurations emphasizing either global or local branches underperform, confirming the necessity of combining global context with local detail. Increased uncertainty weighting (C4) improves COVID detection (82.43\% accuracy, 80.57\% F1) but slightly reduces performance on Lung and Kidney datasets where target features are more prominent. C5 (Consistency-Driven) excels on the Lung dataset (97.86\% accuracy) where structural patterns are clearer, while uniform scaling of all components (C9 \& C10) shows no improvement, indicating that relative balance matters more than absolute weight values.

\subsubsection{Patch Extraction Parameters}

Table~\ref{tab:patch_params} presents F1 scores for different combinations of patch sizes and counts across datasets. Optimal configurations vary by task: kidney abnormality detection performs best with 64×64 patches and 3 patches per image (F1 = 0.9945), lung cancer classification with 64×64 and 2 patches (F1 = 0.9764), and COVID-19 detection with 64×64 and 4 patches (F1 = 0.7903). A patch size of 64 consistently outperforms both smaller (32) and larger (96) sizes, suggesting it provides an optimal balance between local detail and contextual information. The number of required patches aligns with each task's complexity - COVID detection needs more regions due to diffuse disease manifestations, while lung cancer classification can focus on fewer, more localized abnormalities.

\section{Conclusion}
\label{sec:conclusion}

This paper proposed UGPL (Uncertainty-Guided Progressive Learning), a framework for medical image classification that analyzes CT images in two stages: global prediction with uncertainty estimation, followed by local refinement on selected high-uncertainty regions. Our evidential learning-based uncertainty estimation identifies diagnostically challenging areas, while the adaptive fusion mechanism combines global and local predictions based on confidence measures. Extensive experiments across three diverse CT classification tasks (COVID-19 detection, lung cancer classification, and kidney abnormality identification) demonstrate that UGPL significantly outperforms existing methods. Ablations show that the uncertainty-guided patch selection yields upto $5.3 \times$ F1 improvement compared to other configurations. Future work will explore extending UGPL to other modalities like MRI/PET and investigating its potential for uncertainty-guided active learning.

{
    \small
    \bibliographystyle{ieeenat_fullname}
    \bibliography{main}
}

\clearpage
\setcounter{page}{1}
\maketitlesupplementary
\renewcommand{\thesection}{A\arabic{section}}
\setcounter{section}{0}

\section{Extended Methodology}
\label{sec:extended_method}

\subsection{Global Uncertainty Estimation and Evidential Learning}
\label{subsec:supp_global_uncertainty}

Our global uncertainty estimator serves two critical functions: producing initial class predictions and generating a spatial uncertainty map to guide subsequent patch selection. We formulate this as an evidential learning problem that explicitly models uncertainty in the classification process.

\subsubsection{Global Model Architecture}

Given an input CT image $\mathbf{I} \in \mathbb{R}^{H \times W \times 1}$, we employ a ResNet backbone~\cite{resnet} $\mathcal{F}_\theta$ with parameters $\theta$ to extract feature maps $\mathbf{F} \in \mathbb{R}^{h \times w \times d}$, where $h = H/32$, $w = W/32$, and $d$ is the feature dimension:

To accommodate grayscale CT images, we modify the first convolutional layer of the ResNet~\cite{resnet} to accept single-channel inputs while preserving the pretrained weights by averaging across the RGB channels. The feature maps $\mathbf{F}$ are then processed by two parallel heads: a classification head $\mathcal{C}_\phi$ and an evidence head $\mathcal{E}_\psi$. The classification head applies global average pooling followed by a fully connected layer to produce class logits:

\begin{equation}
\mathbf{z}_g = \mathcal{C}_\phi(\mathbf{F}) = \mathbf{W}_\phi \cdot \text{GAP}(\mathbf{F}) + \mathbf{b}_\phi
\end{equation}

where $\mathbf{z}_g \in \mathbb{R}^C$ represents the logits for $C$ classes, $\mathbf{W}_\phi \in \mathbb{R}^{C \times d}$ and $\mathbf{b}_\phi \in \mathbb{R}^C$ are learnable parameters, and $\text{GAP}$ denotes global average pooling.

\begin{equation}
\mathbf{F} = \mathcal{F}_\theta(\mathbf{I})
\end{equation}

\begin{algorithm}[t]
\caption{Global Uncertainty Estimation}
\label{alg:uncertainty}
\begin{algorithmic}[1]
\REQUIRE Input image $\mathbf{I} \in \mathbb{R}^{H \times W \times 1}$
\ENSURE Global logits $\mathbf{z}_g$, Uncertainty map $\hat{\mathbf{U}}$

\STATE $\mathbf{F} \leftarrow \mathcal{F}_\theta(\mathbf{I})$ \COMMENT{Extract features using backbone}
\STATE $\mathbf{z}_g \leftarrow \mathcal{C}_\phi(\mathbf{F})$ \COMMENT{Compute global logits}
\STATE $\mathbf{E} \leftarrow \mathcal{E}_\psi(\mathbf{F})$ \COMMENT{Generate evidence parameters}

\FOR{each spatial location $(i,j)$ and class $c$}
   \STATE $\beta_{i,j,c} \leftarrow \text{softplus}(\mathbf{E}_{i,j,c}) + \epsilon$
   \STATE $\nu_{i,j,c} \leftarrow \frac{e^{\mathbf{E}_{i,j,c+C}}}{\sum_{k=1}^{C} e^{\mathbf{E}_{i,j,k+C}}}$
   \STATE $\alpha_{i,j,c} \leftarrow \beta_{i,j,c} \cdot \nu_{i,j,c} + 1$
\ENDFOR

\FOR{each spatial location $(i,j)$}
   \STATE $\mathbf{U}_{i,j} \leftarrow \frac{1}{C} \sum_{c=1}^{C} \left(\frac{1}{\alpha_{i,j,c}} + \frac{\beta_{i,j,c}}{\alpha_{i,j,c}(\alpha_{i,j,c}+1)}\right)$
\ENDFOR

\STATE $\hat{\mathbf{U}} \leftarrow \frac{\mathbf{U} - \min(\mathbf{U})}{\max(\mathbf{U}) - \min(\mathbf{U}) + \epsilon}$ \COMMENT{Normalize uncertainty map}

\RETURN $\mathbf{z}_g$, $\hat{\mathbf{U}}$
\end{algorithmic}
\end{algorithm}

\subsubsection{Evidential Uncertainty Estimation}

The evidence head $\mathcal{E}_\psi$ generates pixel-wise Dirichlet concentration parameters that quantify uncertainty at each spatial location:

\begin{equation}
\mathbf{E} = \mathcal{E}_\psi(\mathbf{F}) \in \mathbb{R}^{h \times w \times 4C}
\end{equation}

Here, $\mathbf{E}$ encodes four parameters $(\alpha, \beta, \gamma, \nu)$ for each class at each spatial location, representing a Dirichlet distribution. We implement $\mathcal{E}_\psi$ as a sequence of convolutional layers that preserve spatial dimensions while expanding the channel dimension to $4C$. Following the principles of subjective logic \cite{josang2016subjective}, we parameterize the Dirichlet distribution using these four parameters:

\begin{equation}
\alpha_{i,j,c} = \beta_{i,j,c} \cdot \nu_{i,j,c} + 1
\end{equation}

where $(i,j)$ denotes spatial location, $c$ indicates the class, and $\alpha_{i,j,c} > 0$ is the concentration parameter for class $c$ at location $(i,j)$. The parameters $\beta_{i,j,c} > 0$ represents the inverse of uncertainty, $\nu_{i,j,c}$ represents the mass belief, and we constrain $\sum_{c=1}^{C} \nu_{i,j,c} = 1$ to ensure the mass beliefs form a valid probability distribution.

To ensure numerical stability, we apply a softplus activation $f(x) = \log(1 + e^x)$ to compute $\beta_{i,j,c}$ and a softmax function across the class dimension to compute $\nu_{i,j,c}$:

\begin{equation}
\beta_{i,j,c} = f(\mathbf{E}_{i,j,c}) + \epsilon
\end{equation}

\begin{equation}
\nu_{i,j,c} = \frac{e^{\mathbf{E}_{i,j,c+C}}}{\sum_{k=1}^{C} e^{\mathbf{E}_{i,j,k+C}}}
\end{equation}

where $\epsilon$ is a small positive constant for numerical stability. From these parameters, we compute the pixel-wise uncertainty map $\mathbf{U} \in \mathbb{R}^{h \times w}$ by aggregating the uncertainty across all classes:

\begin{equation}
\mathbf{U}_{i,j} = \frac{1}{C} \sum_{c=1}^{C} \left(\frac{1}{\alpha_{i,j,c}} + \frac{\beta_{i,j,c}}{\alpha_{i,j,c}(\alpha_{i,j,c}+1)}\right)
\end{equation}

This formulation captures both aleatoric uncertainty (first term) and epistemic uncertainty (second term). The aleatoric component $\frac{1}{\alpha_{i,j,c}}$ represents uncertainty due to inherent data noise, while the epistemic component $\frac{\beta_{i,j,c}}{\alpha_{i,j,c}(\alpha_{i,j,c}+1)}$ represents uncertainty due to model knowledge limitations.

We normalize the uncertainty map to the range $[0,1]$ for easier interpretation and subsequent processing:

\begin{equation}
\hat{\mathbf{U}} = \frac{\mathbf{U} - \min(\mathbf{U})}{\max(\mathbf{U}) - \min(\mathbf{U}) + \epsilon}
\end{equation}

This normalized uncertainty map $\hat{\mathbf{U}}$ is then used to guide the patch selection process, focusing attention on regions where the global model exhibits high uncertainty. Algorithm~\ref{alg:uncertainty} summarizes the complete process for generating the global class predictions and uncertainty map. The uncertainty map $\hat{\mathbf{U}}$ provides spatial localization of regions where the global model is uncertain about its predictions. High values in $\hat{\mathbf{U}}$ indicate regions that require further analysis through local patch examination. This uncertainty-guided approach allows our model to focus computational resources on diagnostically relevant regions.

\subsection{Uncertainty-Guided Patch Selection and Local Refinement}
\label{subsec:supp_patch_selection}

\subsubsection{Progressive Patch Extraction}

Given an input image $\mathbf{I} \in \mathbb{R}^{H \times W \times 1}$ and its corresponding uncertainty map $\hat{\mathbf{U}} \in \mathbb{R}^{h \times w}$ from the global model, we first upsample the uncertainty map to match the input resolution:

\begin{equation}
\mathbf{U}' = \mathcal{U}(\hat{\mathbf{U}}, (H, W))
\end{equation}

where $\mathcal{U}$ represents bilinear upsampling to dimensions $(H, W)$. Our objective is to extract $K$ patches of size $P \times P$ from regions with high uncertainty while ensuring diversity among the selected patches. We formulate this as a sequential optimization problem where each patch is selected to maximize uncertainty while maintaining a minimum distance from previously selected patches. For the first patch, we simply select the region with maximum uncertainty:

\begin{equation}
(x_1, y_1) = \arg\max_{(x,y)} \mathbf{U}'_{x:x+P, y:y+P}
\end{equation}

where $(x_1, y_1)$ represents the top-left corner of the first patch, and $\mathbf{U}'_{x:x+P, y:y+P}$ denotes the mean uncertainty within the patch region. For subsequent patches $k = 2, 3, \ldots, K$, we introduce a spatial penalty to encourage diversity:

\begin{align}
(x_k, y_k) = \arg\max_{(x,y)} \Big[\, 
    & \mathbf{U}'_{x:x+P,\, y:y+P} \notag \\
    & - \lambda \cdot \min_{i<k} d((x,y),\, (x_i, y_i)) 
\,\Big]
\end{align}

where $d((x,y), (x_i, y_i))$ computes the Euclidean distance between patch centers, $\lambda$ is a weighting parameter controlling diversity, and $\min_{i<k}$ finds the minimum distance to any previously selected patch. To implement this efficiently while avoiding explicit computation of the penalty term for all possible locations, we apply a non-maximum suppression (NMS) approach. After selecting each patch, we suppress a region around it by applying a penalty mask to the uncertainty map:

\begin{align}
\mathbf{U}'_{x\!-\!M:x\!+\!P\!+\!M,\, y\!-\!M:y\!+\!P\!+\!M}
=\, & \mathbf{U}'_{x\!-\!M:x\!+\!P\!+\!M,\, y\!-\!M:y\!+\!P\!+\!M} \notag \\
& \times \left( 1 - \mathbf{G} \right)
\end{align}

where $M$ is a margin parameter and $\mathbf{G}$ is a Gaussian kernel that applies a stronger suppression near the center of the selected patch and gradually reduces toward the edges.

Our algorithm incorporates several fallback mechanisms to handle edge cases and ensure reliable operation:

\begin{itemize}
    \item \textbf{Uncertainty Threshold Handling:} In situations where no high-uncertainty regions remain (when all uncertainty values are suppressed below a specified threshold), the method falls back to random selection to preserve sample diversity.

    \item \textbf{Boundary Checking:} Comprehensive boundary checking is applied to prevent selected patches from extending beyond the image borders, ensuring valid patch extraction even at image edges.

    \item \textbf{Dynamic Size Adjustment:} To accommodate extremely small images or atypical aspect ratios, the algorithm dynamically adjusts patch sizes, ensuring consistent and valid outputs across varying input dimensions.
\end{itemize}

These mechanisms collectively ensure robust operation across diverse medical imaging datasets with varying characteristics.

\subsubsection{Local Refinement Network Architecture}

After extracting the $K$ patches $\{\mathbf{P}_1, \mathbf{P}_2, \ldots, \mathbf{P}_K\}$, we process each patch independently using a local refinement network. This network comprises three components: a feature extractor, a classification head, and a confidence estimation head. 

The feature extractor $\mathcal{L}_f$ processes each patch to obtain local feature vectors:
\begin{equation}
\mathbf{f}_k = \mathcal{L}_f(\mathbf{P}_k) \in \mathbb{R}^{d_l}
\end{equation}
where $d_l$ is the feature dimension. We implement $\mathcal{L}_f$ as a sequence of convolutional layers followed by pooling operations to progressively reduce spatial dimensions while increasing feature depth. Specifically, our implementation uses four convolutional blocks with increasing channel dimensions (64→128→256→256), each followed by batch normalization, ReLU activation, and max pooling. The final features undergo adaptive average pooling to produce a fixed-dimensional representation regardless of input patch size.

The classification head $\mathcal{L}_c$ maps these features to class logits:

\begin{equation}
\mathbf{z}_{l,k} = \mathcal{L}_c(\mathbf{f}_k) \in \mathbb{R}^C
\end{equation}

This head is implemented as a two-layer MLP with a hidden dimension of 128 and ReLU activation between layers. Simultaneously, the confidence estimation head $\mathcal{L}_\text{conf}$ produces a scalar confidence score for each patch:

\begin{equation}
c_k = \mathcal{L}_\text{conf}(\mathbf{f}_k) \in [0, 1]
\end{equation}

where $c_k$ represents the model's confidence in its prediction for patch $k$. We implement $\mathcal{L}_\text{conf}$ as a small MLP with a sigmoid activation function on the output to constrain the confidence score to the range $[0, 1]$. This two-layer MLP has a hidden dimension of 64 and uses ReLU activation between layers.

The confidence score serves two critical purposes: (1) it allows the model to express uncertainty about individual patch predictions, and (2) it provides a weight for the subsequent fusion of local predictions. Patches with higher confidence scores will contribute more significantly to the final classification decision. For each patch $k$, we obtain both class logits $\mathbf{z}_{l,k}$ and a confidence score $c_k$. The combined local prediction is computed as a confidence-weighted average of the patch predictions:

\begin{equation}
\mathbf{z}_l = \frac{\sum_{k=1}^K c_k \cdot \mathbf{z}_{l,k}}{\sum_{k=1}^K c_k + \epsilon}
\end{equation}

where $\epsilon$ is a small constant (typically $10^{-6}$) for numerical stability. This formulation naturally handles cases where some patches have very low confidence, effectively reducing their contribution to the final prediction.

The local refinement network provides detailed analysis of suspicious regions identified by the global model, capturing fine-grained features that might be missed in the global analysis. By assigning confidence scores to each patch, the network also performs an implicit form of attention, focusing on the most discriminative patches for the final classification decision.

\subsection{Adaptive Fusion and Training Objectives}
\label{subsec:supp_fusion_training}

\subsubsection{Adaptive Fusion Module}

The adaptive fusion module dynamically determines the optimal weighting between global and local predictions for each input image. Given the global logits $\mathbf{z}_g \in \mathbb{R}^C$ and uncertainty map $\hat{\mathbf{U}} \in \mathbb{R}^{h \times w}$ from the global model, and local logits $\mathbf{z}_l \in \mathbb{R}^C$ with patch confidence scores $\{c_1, c_2, \ldots, c_K\}$ from the local refinement network, we compute a scalar representation of the global uncertainty by averaging across the spatial dimensions:

\begin{equation}
u_g = \frac{1}{h \cdot w} \sum_{i=1}^h \sum_{j=1}^w \hat{\mathbf{U}}_{i,j}
\end{equation}

This scalar uncertainty $u_g \in [0, 1]$ quantifies the overall confidence of the global model. The fusion network $\mathcal{F}_\omega$ takes as input the global logits $\mathbf{z}_g$ and the global uncertainty score $u_g$, concatenated into a single vector $[\mathbf{z}_g, u_g] \in \mathbb{R}^{C+1}$. The network outputs a fusion weight $w_g \in [0, 1]$ that determines the relative contribution of global versus local predictions:

\begin{equation}
w_g = \mathcal{F}_\omega([\mathbf{z}_g, u_g])
\end{equation}

We implement $\mathcal{F}_\omega$ as a multi-layer perceptron with sigmoid activation on the output:

\begin{equation}
\mathcal{F}_\omega([\mathbf{z}_g, u_g]) = \sigma(W_2 \cdot \text{ReLU}(W_1 \cdot [\mathbf{z}_g, u_g] + b_1) + b_2)
\end{equation}

where $W_1 \in \mathbb{R}^{d_f \times (C+1)}$, $W_2 \in \mathbb{R}^{1 \times d_f}$, $b_1 \in \mathbb{R}^{d_f}$, and $b_2 \in \mathbb{R}$ are learnable parameters, $d_f$ is the hidden dimension, and $\sigma$ is the sigmoid function. The fusion weight $w_g$ represents the contribution of the global prediction, while $w_l = 1 - w_g$ represents the contribution of the local prediction. The fused logits $\mathbf{z}_f$ are computed as:

\begin{equation}
\mathbf{z}_f = w_g \cdot \mathbf{z}_g + (1 - w_g) \cdot \mathbf{z}_l
\end{equation}

This adaptive weighting allows the model to rely more on global features when the global model is confident (low uncertainty), and more on local features when the global model is uncertain (high uncertainty).

\begin{figure*}[t]
    \centering
    \subfloat[Global feature embeddings showing clear class separation with distinct clusters for each kidney condition (Normal, Cyst, Tumor, Stone). The global model learns discriminative whole-image representations that establish strong decision boundaries between classes.]{\includegraphics[width=0.48\textwidth]{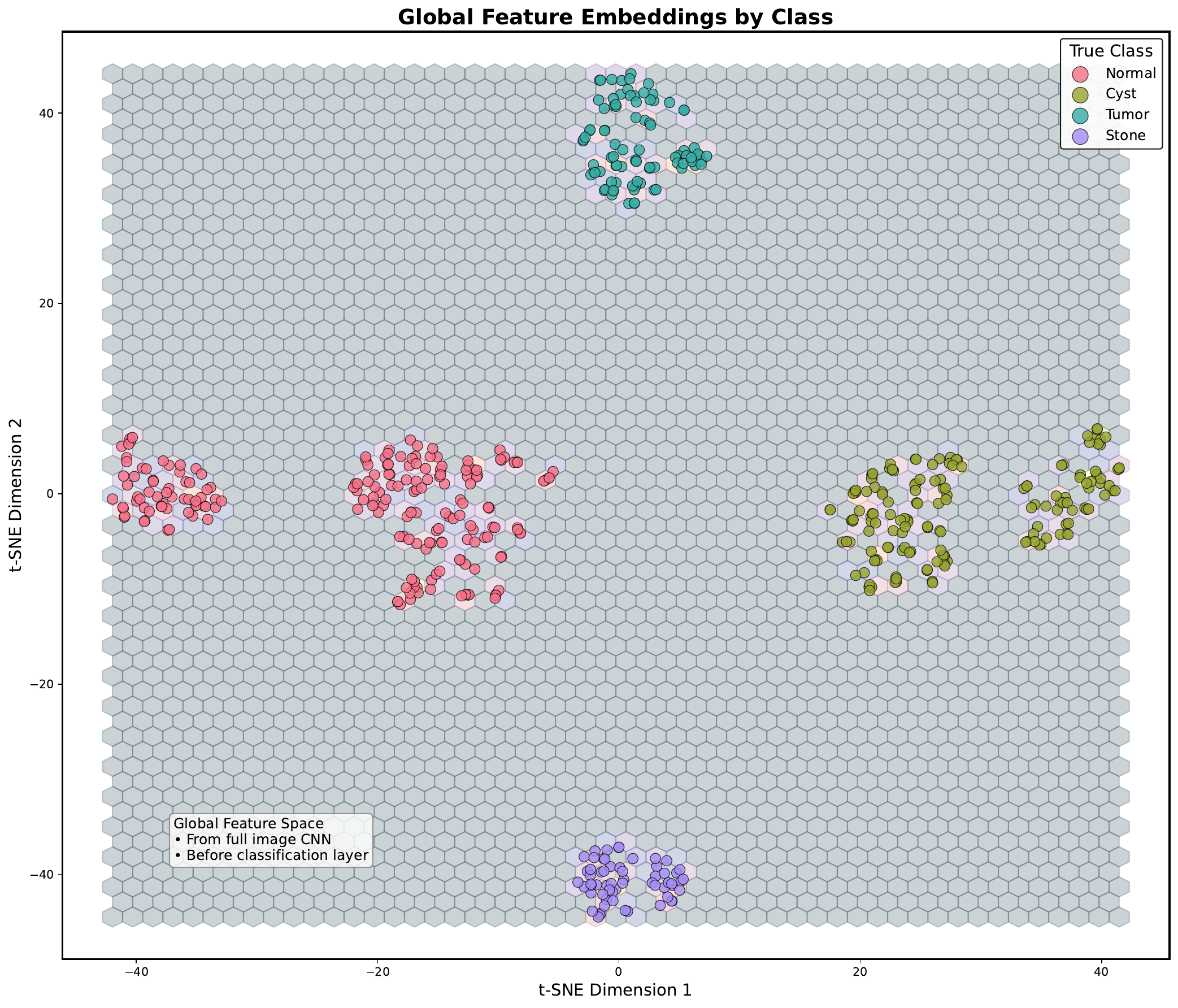}\label{fig:global_embeddings}}
    \hfill
    \subfloat[Local feature embeddings exhibiting significant class mixing without distinct clusters. The local model focuses on fine-grained details within uncertain regions, capturing complementary information not directly aligned with class boundaries.]{\includegraphics[width=0.48\textwidth]{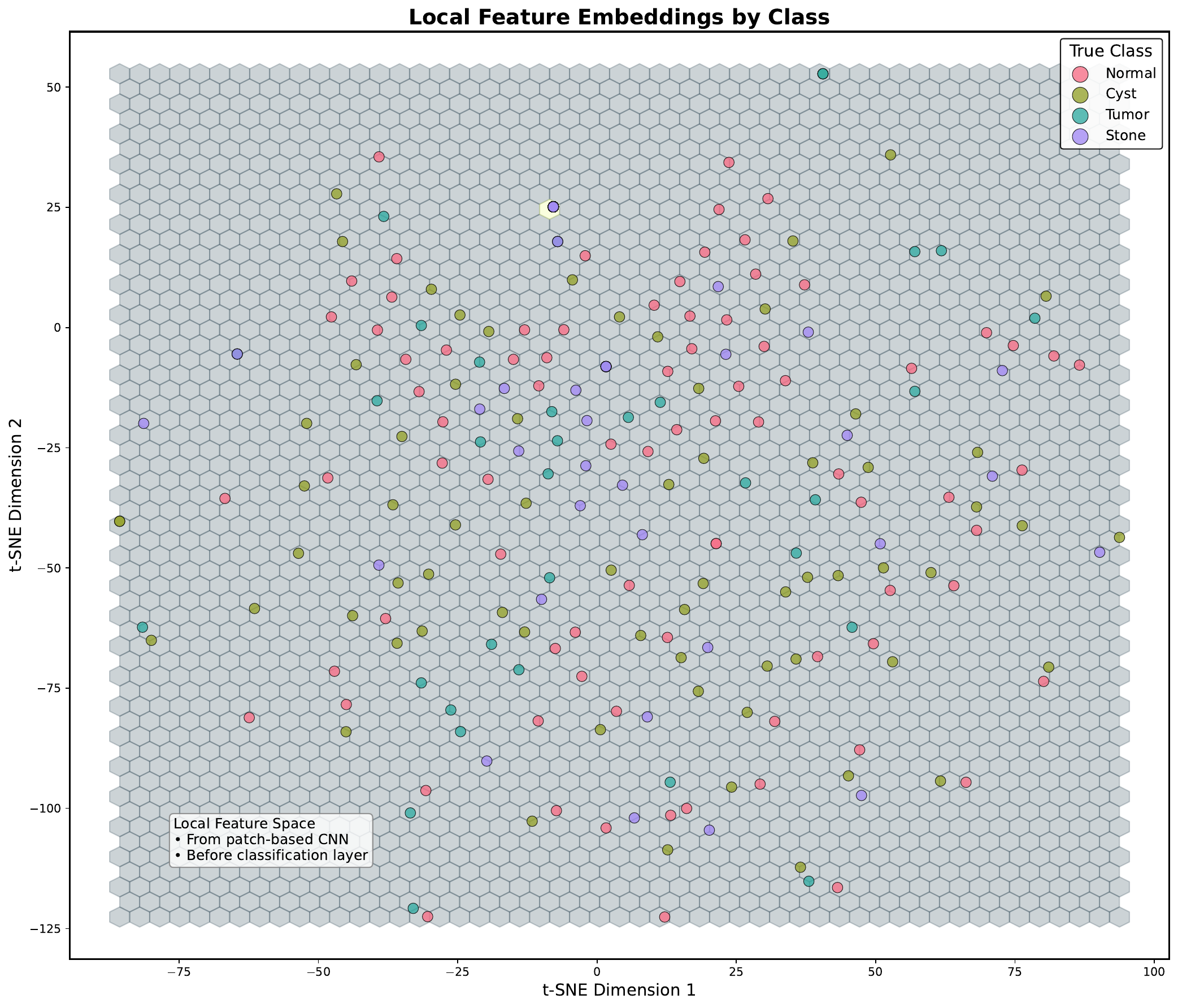}\label{fig:local_embeddings}}
    
    \caption{Comparison of t-SNE visualizations for feature spaces in the kidney dataset. (a) Global features from the full-image CNN form well-separated clusters by class, demonstrating effective overall classification capability. (b) Local features from patch-based analysis show substantial mixing across classes, indicating their focus on subtle variations and uncertainty resolution rather than direct class discrimination. This complementary representation underscores why adaptive fusion of both feature types produces superior performance.}
    \label{fig:embeddings}
\end{figure*}

\subsubsection{Multi-component Loss Function}

Our comprehensive loss function addresses multiple objectives simultaneously. The total loss $\mathcal{L}_\text{total}$ is a weighted sum of several components:

\begin{align}
\mathcal{L}_\text{total} =\ & \lambda_f \mathcal{L}_\text{fused}
+ \lambda_g \mathcal{L}_\text{global}
+ \lambda_l \mathcal{L}_\text{local} \nonumber \\
& + \lambda_u \mathcal{L}_\text{uncertainty}
+ \lambda_c \mathcal{L}_\text{consistency}
+ \lambda_{\text{conf}} \mathcal{L}_\text{confidence} \nonumber \\
& + \lambda_d \mathcal{L}_\text{diversity}
\end{align}

where $\lambda_f, \lambda_g, \lambda_l, \lambda_u, \lambda_c, \lambda_\text{conf}$, and $\lambda_d$ are weighting coefficients for each loss component.

\paragraph{Classification Losses.} We apply cross-entropy loss to the predictions from each component of our framework:

\begin{equation}
\mathcal{L}_\text{fused} = -\sum_{i=1}^C y_i \log(\text{softmax}(\mathbf{z}_f)_i)
\end{equation}

\begin{equation}
\mathcal{L}_\text{global} = -\sum_{i=1}^C y_i \log(\text{softmax}(\mathbf{z}_g)_i)
\end{equation}

\begin{equation}
\mathcal{L}_\text{local} = \frac{1}{K} \sum_{k=1}^K -\sum_{i=1}^C y_i \log(\text{softmax}(\mathbf{z}_{l,k})_i)
\end{equation}

where $y_i$ is the ground truth label for class $i$ (one-hot encoded), and $\text{softmax}(\mathbf{z})_i$ denotes the softmax probability for class $i$ given logits $\mathbf{z}$.

\paragraph{Uncertainty Calibration Loss.} To ensure that the uncertainty map accurately reflects prediction errors, we introduce an uncertainty calibration loss:

\begin{equation}
\mathcal{L}_\text{uncertainty} = \text{MSE}(\hat{\mathbf{U}}, 1 - \mathbf{C})
\end{equation}

where $\mathbf{C} \in \{0, 1\}^{h \times w}$ is a correctness map derived from the global predictions. For each spatial location $(i,j)$, $\mathbf{C}_{i,j} = 1$ if the predicted class at that location matches the ground truth, and $\mathbf{C}_{i,j} = 0$ otherwise. This loss encourages high uncertainty in regions where the global model makes errors and low uncertainty where predictions are correct.

\paragraph{Consistency Loss.} To promote consistency between global and local predictions, we use a Kullback-Leibler (KL) divergence loss:

\begin{equation}
\mathcal{L}_\text{consistency} = \frac{1}{K} \sum_{k=1}^K \text{KL}(\text{softmax}(\mathbf{z}_{l,k}) \| \text{softmax}(\mathbf{z}_g)) \cdot c_k
\end{equation}

where $\text{KL}(P \| Q) = \sum_i P_i \log(P_i / Q_i)$ is the KL divergence, and $c_k$ is the confidence score for patch $k$. This loss is weighted by the patch confidence, reducing the penalty for inconsistency in low-confidence patches.

\paragraph{Confidence Regularization Loss.} To align patch confidence scores with prediction accuracy, we introduce a confidence regularization loss:

\begin{equation}
\mathcal{L}_\text{confidence} = \frac{1}{K} \sum_{k=1}^K \text{MSE}(c_k, a_k)
\end{equation}

where $a_k \in \{0, 1\}$ indicates whether the prediction for patch $k$ is correct ($a_k = 1$) or incorrect ($a_k = 0$). This loss encourages high confidence for correct predictions and low confidence for incorrect predictions.

\paragraph{Diversity Loss.} To encourage diversity among patch predictions, we include a diversity loss:

\begin{align}
\mathcal{L}_\text{diversity} =\ 
& \frac{1}{K(K-1)/2} \sum_{i=1}^{K-1} \sum_{j=i+1}^K \nonumber \\
& \text{cos}\left(\text{softmax}(\mathbf{z}_{l,i}),\, \text{softmax}(\mathbf{z}_{l,j})\right)
\end{align}

where $\text{cos}(a, b) = \frac{a \cdot b}{||a|| \cdot ||b||}$ is the cosine similarity between vectors. This loss penalizes similarity between patch predictions, encouraging each patch to contribute unique information.

\begin{figure}[t]
    \centering
    \includegraphics[width=\columnwidth]{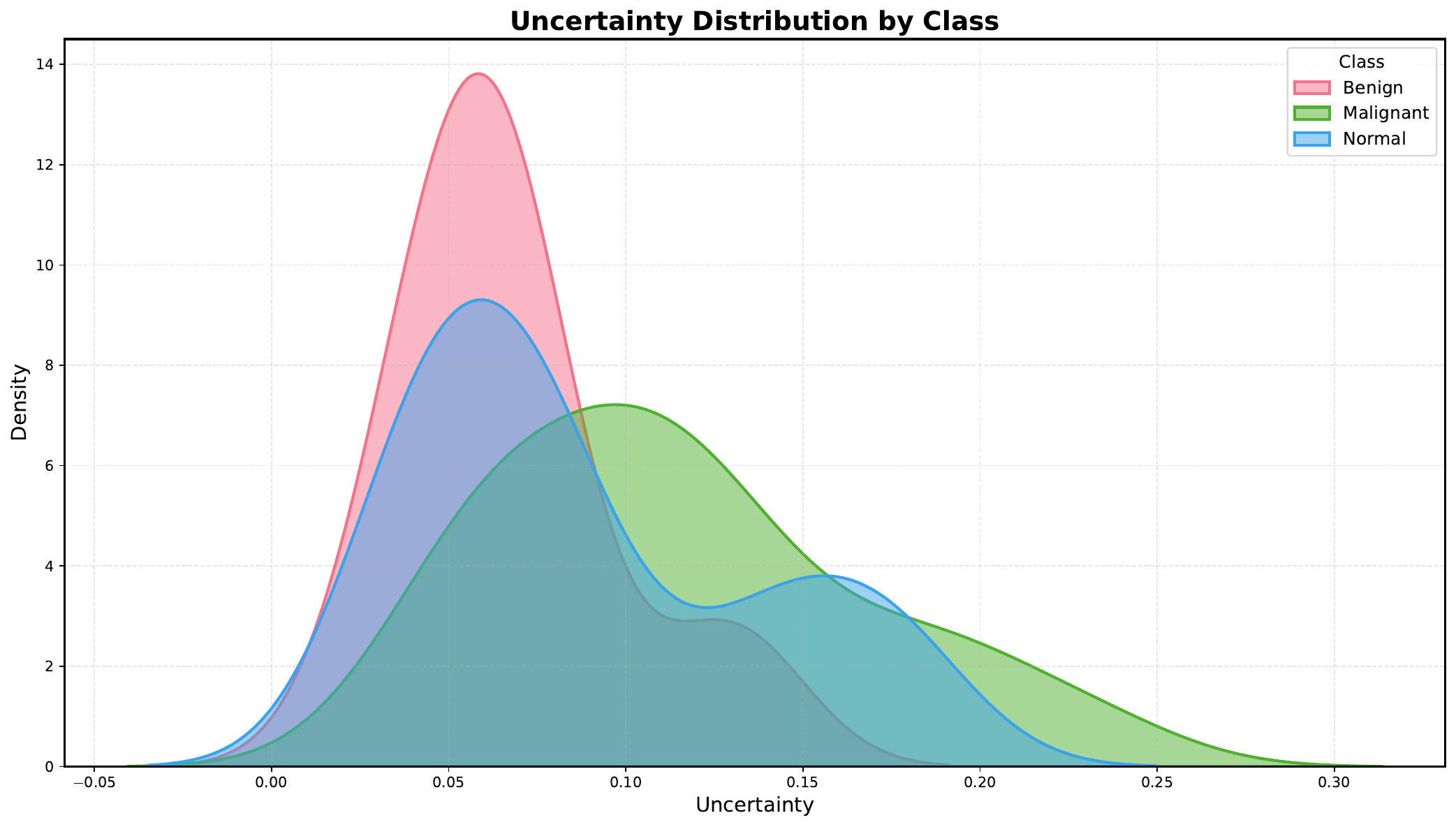}
    \caption{Uncertainty distribution by class for lung cancer detection. Malignant cases (green) exhibit significantly higher average uncertainty and broader distribution compared to benign cases (pink), which show a tighter, lower-uncertainty distribution. Normal cases (blue) display a distinctive bimodal distribution with peaks at both low and moderate uncertainty levels.}
    \label{fig:uncertainty_dist}
\end{figure}

\section{Implementation Details}
\label{sec:impl_details}

\begin{figure}[t]
    \centering
    \includegraphics[width=\columnwidth]{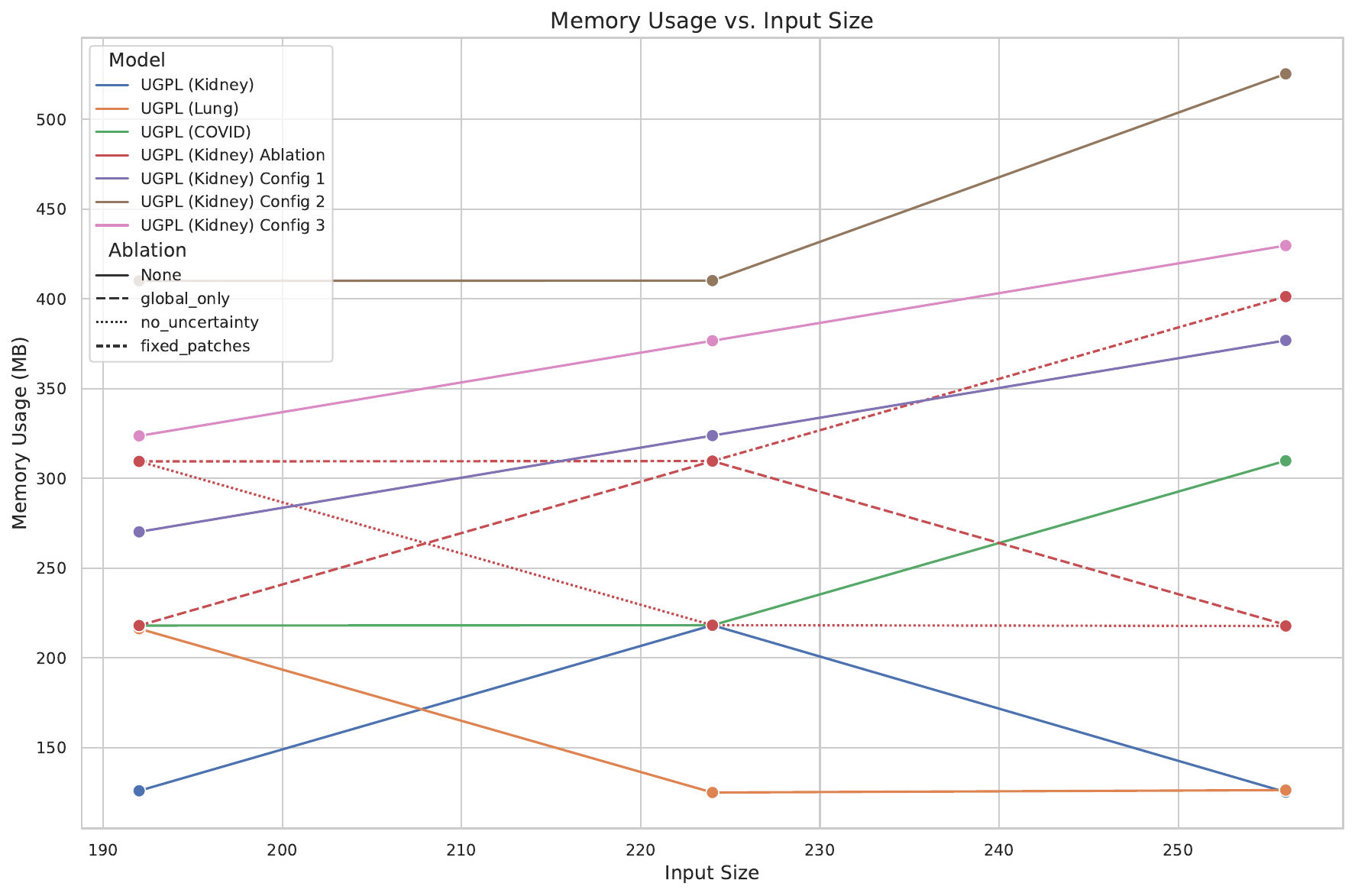}
    \caption{Memory usage scaling with input dimensions across UGPL variants. Lines represent different model configurations and ablations. Config 2 (brown line) consistently demonstrates the highest memory requirements due to its ResNet-50~\cite{resnet} backbone variant. Some configurations show counterintuitive scaling behavior, particularly at larger input sizes, highlighting complex interactions between model architecture and GPU memory management.}
    \label{fig:memory_usage}
\end{figure}

All models are trained for 100 epochs with early stopping based on validation loss with a patience of 7 epochs on a single NVIDIA RTX 3090 GPU. We employ an Adam optimizer~\cite{AdamOpt} with a learning rate of $1 \times 10^{-4}$ and weight decay of $1 \times 10^{-4}$, with a batch size of 96 and a cosine decay learning rate scheduler~\cite{scheduler}. For data augmentation~\cite{albumentations} during training, we apply random horizontal and vertical flips, random rotation (±10°), random affine transformations (±5\% translation), and contrast/brightness adjustments (±10\%). Images are normalized to the [0,1] range after applying appropriate windowing for CT images. We do not use EMA~\cite{EMA} since it does not improve performance.

\begin{figure*}[t]
    \centering
    
    \includegraphics[width=\textwidth]{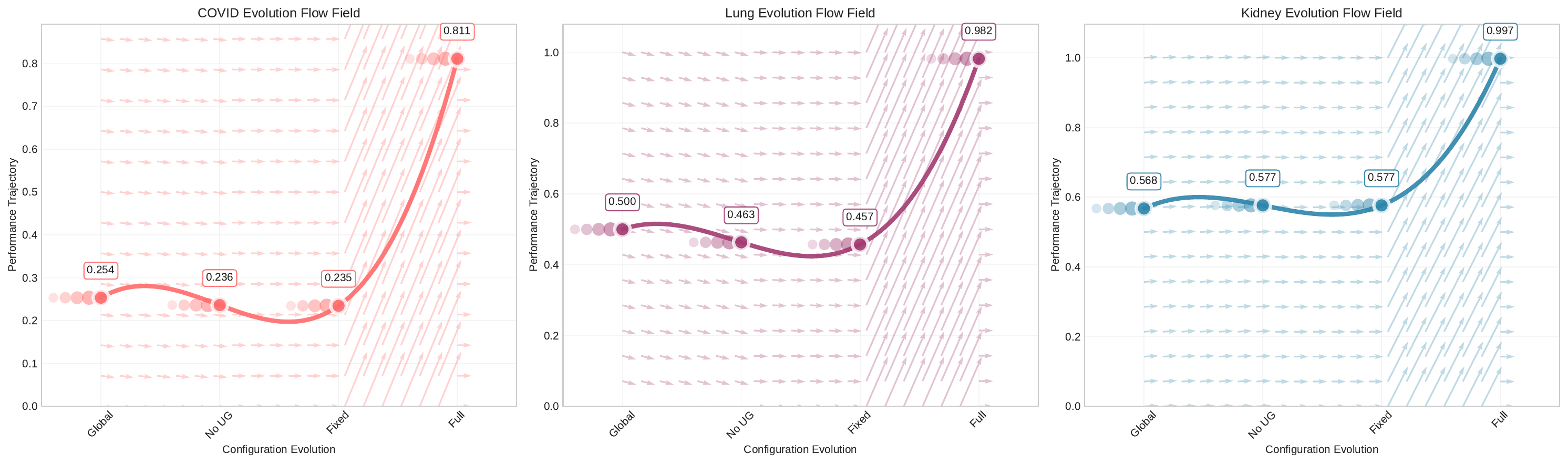}\\[0.5em]
    
    \begin{tabular}{ccc}
        \includegraphics[width=0.3\textwidth]{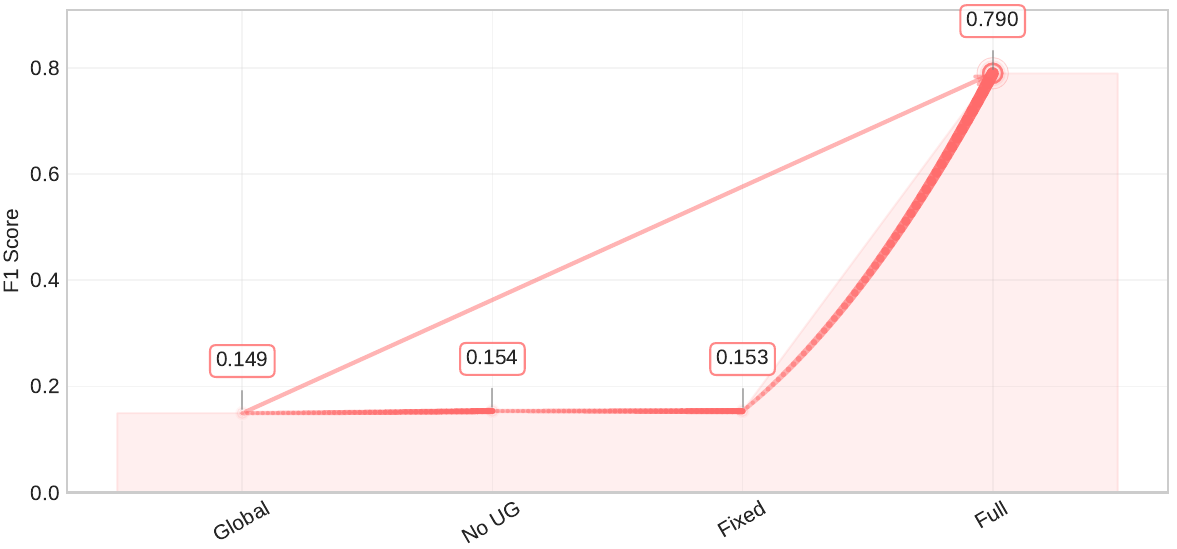} &
        \includegraphics[width=0.3\textwidth]{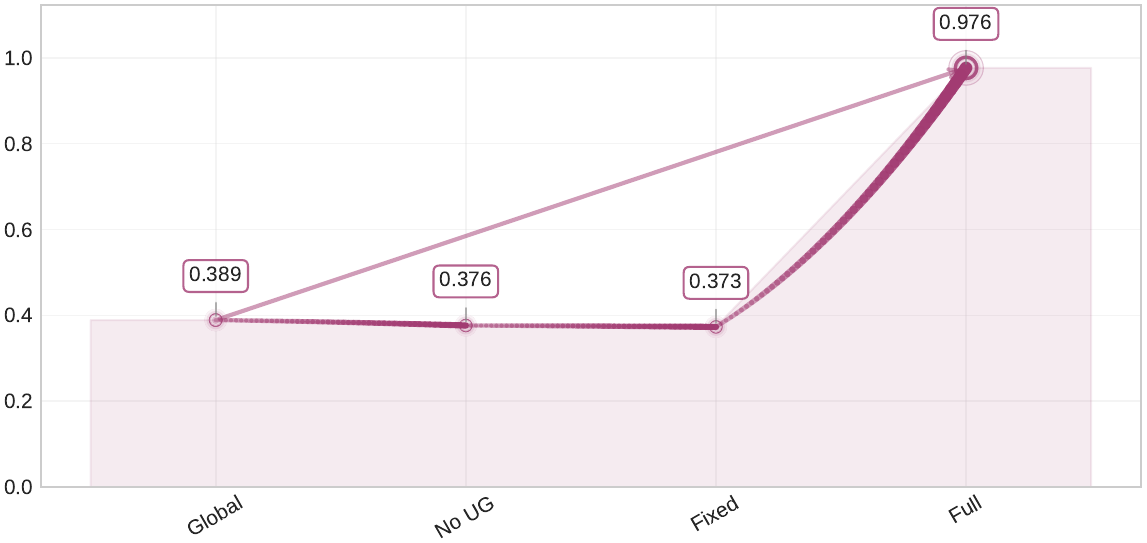} &
        \includegraphics[width=0.3\textwidth]{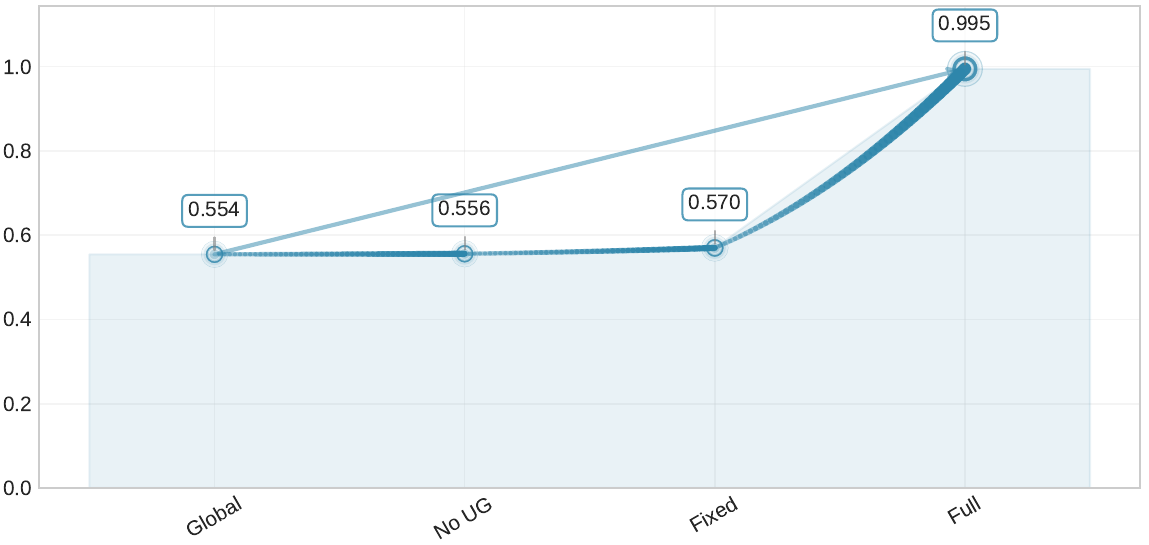}
    \end{tabular}
    
    \caption{Evolution of model performance across different configurations. Top: Flow field visualization showing performance trajectories from simplified to complete model configurations for each dataset. Bottom: F1 score progression across configurations for COVID (left), Lung (middle), and Kidney (right) datasets, highlighting the dramatic improvement when all components are integrated in the full model.}
    \label{fig:ablation_evolution}
\end{figure*}

Model configurations are adapted for each dataset as follows: the Kidney dataset uses a ResNet-18~\cite{resnet} backbone with a patch size of 64 and 3 patches per image, the Lung dataset uses a ResNet-50~\cite{resnet} backbone with a patch size of 64 and 2 patches per image, and the COVID dataset uses a ResNet-18~\cite{resnet} backbone with a patch size of 64 and 4 patches per image. The multi-component loss function assigns weights of 1.0 for the fused loss, 0.5 for global and local losses, 0.3 for the uncertainty loss, 0.2 for the consistency loss, and 0.1 for both the confidence and diversity losses.

\section{Additional Experiments and Results}
\label{sec:add_results}

\subsection{Feature Space Analysis}
To better understand how UGPL learns different representations at global and local scales, we visualize the feature embeddings from both network components using t-SNE. Figure~\ref{fig:embeddings} demonstrates the contrast between global and local feature spaces for the kidney CT dataset~\cite{kidneydataset}.

The global feature embeddings (Figure~\ref{fig:global_embeddings}) display remarkably clear separation between classes, with distinct clusters forming for each pathological condition. This indicates that the global network successfully learns discriminative features that establish strong decision boundaries at the whole-image level. In contrast, the local feature embeddings (Figure~\ref{fig:local_embeddings}) exhibit substantial mixing between classes with no clear cluster formation, suggesting that the local network captures different characteristics altogether.

The global network provides robust overall classification by learning class-separable features, while the local network focuses on fine-grained details within uncertain regions that may not align directly with class boundaries but capture subtle variations critical for resolving ambiguous cases. When these complementary features are combined through our adaptive fusion mechanism, the model effectively leverages both the discriminative power of global features and the detailed analysis of local features, particularly in challenging regions where global analysis alone might be insufficient.

The dispersed nature of local embeddings also validates our patch selection approach - these patches represent precisely those regions where additional analysis is most beneficial, as they contain ambiguous features that the global model finds difficult to classify confidently. This feature space analysis provides concrete evidence for why progressive refinement is more effective than single-pass approaches for medical image classification.

\subsection{Uncertainty Calibration Analysis}

Figure~\ref{fig:uncertainty_dist} visualizes the distribution of pixel-wise uncertainty values across diagnostic classes in the lung cancer dataset~\cite{lungdataset1,lungdataset2,lungdataset3}. The distinct separation between uncertainty profiles demonstrates the model's ability to calibrate uncertainty in a clinically meaningful way. Malignant cases consistently show higher uncertainty (mean 0.14, standard deviation 0.07) compared to benign cases (mean 0.06, standard deviation 0.03), reflecting the inherently more complex and variable presentation of malignant lesions. Normal cases exhibit an intriguing bimodal distribution, suggesting the existence of two distinct subgroups within what radiologists classify as normal tissue. This aligns with clinical practice, where some normal cases closely resemble benign findings (first mode) while others contain subtle variations that warrant closer inspection (second mode). The UGPL framework effectively leverages these uncertainty patterns to guide computational resource allocation, focusing detailed analysis precisely where diagnostic ambiguity is highest.

\subsection{Ablation Evolution}

Figure~\ref{fig:ablation_evolution} visualizes performance evolution across configurations. All datasets show minimal variations among simplified configurations followed by dramatic jumps with the full model - COVID F1 scores improve 5.3× (0.15 to 0.79), lung dataset by 2.6× (0.37 to 0.98), and kidney dataset by 1.7× (0.57 to 0.99).

\subsection{Computational Efficiency Analysis}

\begin{figure}[t]
    \centering
    \includegraphics[width=\columnwidth]{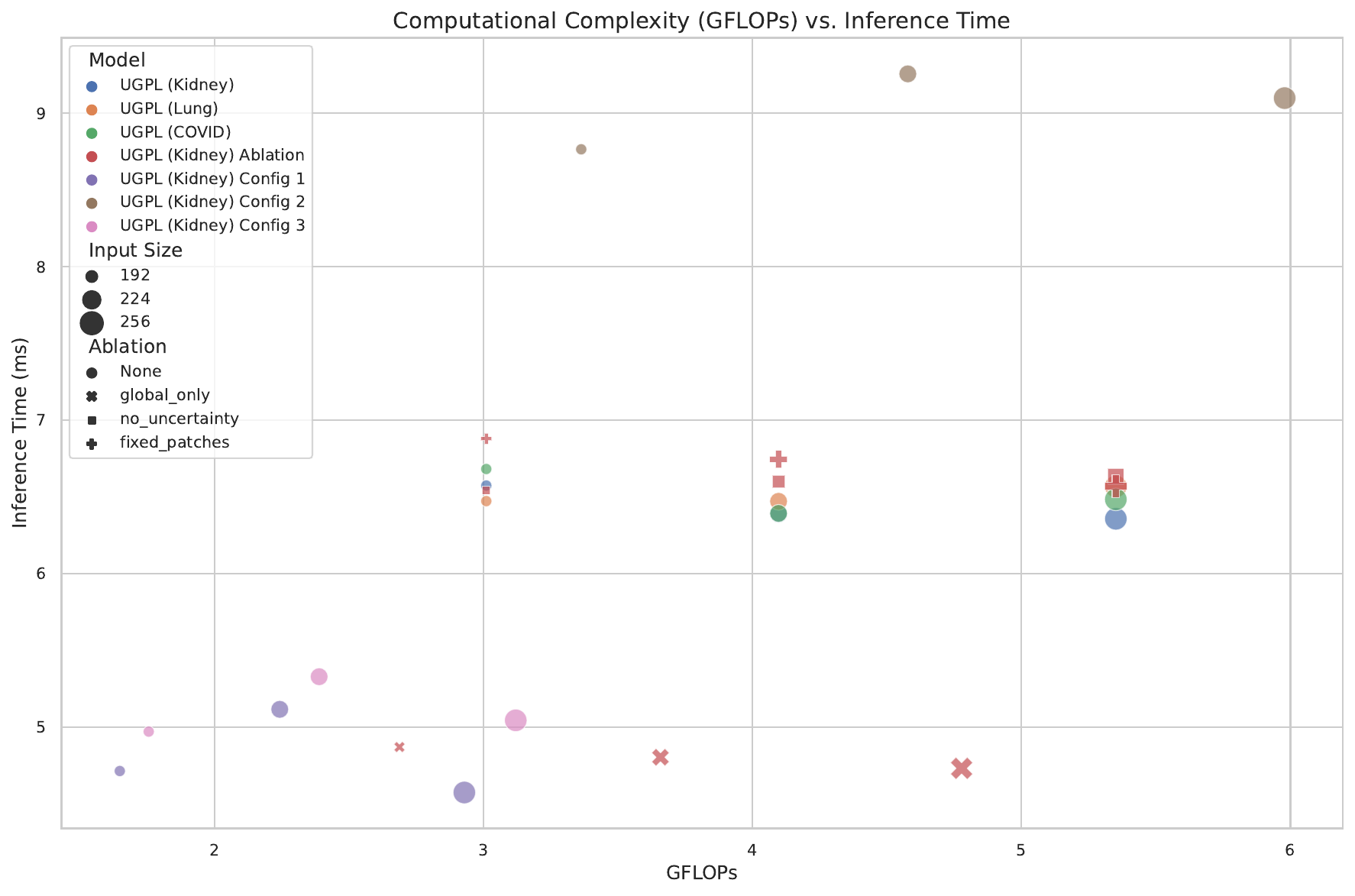}
    \caption{Computational complexity (GFLOPs) versus inference time (ms) for UGPL variants. Points are colored by dataset, with marker style indicating ablation type and size representing input dimensions.}
    \label{fig:comp_efficiency}
\end{figure}

We analyze computational efficiency of UGPL across different configurations and ablations to understand tradeoffs between model complexity and performance. Figure~\ref{fig:comp_efficiency} shows the relationship between computational complexity (measured in GFLOPs) and inference time. The full UGPL model requires approximately 3-5 GFLOPs depending on the dataset and configuration, with inference times between 4.5-6.7ms on an NVIDIA P100 (we use a lightweight GPU for inference to better reflect real-world deployment settings). The global-only ablation (without patch extraction and local refinement) reduces inference time by 27-36\% across all datasets, demonstrating the computational cost of the progressive analysis components. Higher-capacity backbones (Config 2 with ResNet-50 variant) increase both GFLOPs and inference time by approximately 45\% compared to the standard configurations.

Memory efficiency is another critical factor for medical imaging applications. Figure~\ref{fig:memory_usage} illustrates how memory usage scales with input image dimensions. We observe non-linear scaling patterns that vary significantly across configurations. The ResNet-50 backbone (Config 2) requires 1.4-1.7× more memory than ResNet-18 configurations. Interestingly, ablations demonstrate dataset-specific memory profiles: for the COVID dataset, memory usage increases linearly with input size, while the Kidney dataset shows more complex patterns. The global-only ablation demonstrates inconsistent memory scaling, suggesting that optimizations in GPU memory management affect different architectural components differently.

UGPL model requires more computational resources than simplified variants, and the progressive learning approach maintains reasonable efficiency for clinical deployment. The additional cost of uncertainty estimation and local refinement is justified by the significant performance improvements, particularly for challenging cases.

\end{document}